\documentclass[a4paper, amsfonts, amssymb, amsmath, reprint, showkeys, nofootinbib, twoside]{revtex4-1}
\usepackage[english]{babel}
\usepackage[utf8]{inputenc}
\usepackage[colorinlistoftodos, color=green!40, prependcaption]{todonotes}
\usepackage{lineno}
\usepackage{mathrsfs}
\usepackage{amsmath}
\usepackage{xcolor}
\usepackage{graphicx}
\usepackage[title]{appendix}

\DeclareMathOperator{\sech}{sech}
\DeclareMathOperator{\sinc}{sinc}
\usepackage{amsthm}
\usepackage{mathtools}
\usepackage{physics}
\usepackage{xcolor}
\usepackage{graphicx}
\usepackage[left=23mm,right=13mm,top=35mm,columnsep=15pt]{geometry} 
\usepackage{adjustbox}
\usepackage{placeins}
\usepackage[T1]{fontenc}
\usepackage{lipsum}
\usepackage{csquotes}
\usepackage[pdftex, pdftitle={Article}, pdfauthor={Author}]{hyperref} 
\begin{document}
\title{Energy Scalability Limits of Dissipative Solitons}

\author{Vladimir L. Kalashnikov}
    \email[Correspondence email address: ]{vladimir.kalashnikov@ntnu.no}
    \affiliation{Department of Physics, Norwegian University of Science and Technology, 7491 Trondheim, Norway}
\author{Alexander Rudenkov}
    \affiliation{Department of Physics, Norwegian University of Science and Technology, 7491 Trondheim, Norway}
\author{Evgeni Sorokin}
    \affiliation{Institut f{\"u}r Photonik, TU Wien,  Gu{\ss}hausstra{\ss}e 27/387, A-1040 Vienna, Austria}
    \affiliation{ATLA lasers AS, Richard Birkelands vei 2B, 7034 Trondheim, Norway}
\author{Irina T. Sorokina} 
    \affiliation{Department of Physics, Norwegian University of Science and Technology, 7491 Trondheim, Norway}
    \affiliation{ATLA lasers AS, Richard Birkelands vei 2B, 7034 Trondheim, Norway}

\date{\today} 

\begin{abstract}
In this study, we apply a thermodynamical approach to elucidate the primary constraints on the energy scaling of dissipative solitons (DS). We rely on the adiabatic theory of strongly chirped DS and define the DS energy scaling in terms of dissipative soliton resonance (DSR). Three main experimentally verifiable signatures identify a transition to DSR: i) growth of a Lorentzian spike at the centrum of the DS spectrum, which resembles a spectral condensation in Bose-Einstein condensate (BEC), ii) saturation of the spectrum broadening, and iii) asymptotic DS stretching. We connect the DSR breakup with three critical factors: i) decoupling of two correlation scales inherent in strongly chirped DS, ii) resulting rise of the DS entropy with energy, which provokes its disintegration, and iii) transition to a nonequilibrium phase, which is characterized by negative temperature. The breakup results in multiple stable DS generation. Theoretical results are in good qualitative agreement with the experimental data from a Kerr-lens mode-locked Cr$^{2+}$:ZnS chirped-pulse oscillator (CPO) that paves the way for optimizing high-energy femtosecond pulse generation in solid-state CPO and all-normal-dispersion fiber lasers.
\end{abstract}


\maketitle

\section{Introduction}\label{intr}
Recent progress in femtosecond laser technology has catalyzed transformative changes across various scientific and technological domains, including attosecond and relativistic physics, ultrafast spectroscopy, nanophotonics, particle beam acceleration, and material processing \cite{chang2016attosecond}. The deployment of ultrashort light pulses from these lasers has facilitated groundbreaking discoveries in fundamental physics, astrophysics, and quantum electrodynamics, providing new insights into fundamental constants and exploring the depths of relativistic phenomena \cite{mourou2006optics,RevModPhys841177,plaja2013attosecond,guenot2017relativistic,dombi2020strong}. Recently, femtosecond lasers have achieved peak powers at the petawatt (PW) level \cite{danson2019petawatt,yoon2021realization}. 

Specifically, even moderately peak powers yield substantial scientific and industrial applications in ultrashort pulse technology within the mid-infrared wavelength spectrum exceeding 2 microns. These applications include precision material processing of semiconductors and composite materials, and various medical applications \cite{braun2008ultrashort,richter2020sub,ASEC2021-11143}. The systems, compact enough to be accommodated on tabletops and capable of emitting sub- and over-microjoule femtosecond pulses, are exceptionally well-suited for various tasks, including, e.g., environmental monitoring and silicon processing \cite{mirov2014progress,sorokina2014femtosecond}. Looking forward, the development of future compact, high-intensity chirped-pulse oscillator/amplifier (CPO-CPA) systems, which leverage these technologies, is anticipated to deliver significant societal benefits. These benefits include advancements in medical proton therapy, the production of clean nuclear energy, and the transmutation of nuclear waste. The upcoming generation of lasers, notable for their high intensity, compactness, user-friendliness, and megahertz repetition rates, will play a pivotal role in the evolution of these crucial sectors.

The well-established way to effectively scale the energy of ultrashort pulses is to use a mode-locked system with all-normal group-delay dispersion (GDD) in a fiber laser \cite{chong2006all} or solid-state chirped-pulse oscillator (CPO) \cite{fernandez2004chirped}, where the energy harvesting results from a pulse stretching caused by its strong chirp. This approach can generally be considered as applying the dissipative soliton (DS) concept \cite{vanin1994dissipative} to laser mode-locking (ML) \cite{grelu2012dissipative,grelu2024solitary}. Then, the ultrashort pulse energy scaling can be treated as a manifestation of the dissipative soliton resonance (DSR) \cite{PhysRevA.78.023830,grelu2010dissipative}, meaning an asymptotical energy growth just by increasing a laser pulse repetition rate (or, to a lesser extent, by increasing the average power and mode size).

The robustness of DS, allowing energy accumulation in the DSR regime, can be explained by intensive energy exchange with the environment and its redistribution inside the DS \cite{ankiewicz2008dissipative}. The primary mechanism of such redistribution is the phase inhomogeneity of the DS, or chirp \cite{grelu2024solitary}. It creates a nontrivial DS internal structure fraught with coherence loss. Thus, a strongly-chirped DS can be attributed to a class of semi/in-coherent solitons \cite{grelu2024solitary}, which admit the powerful methods of kinetic theory for analysis \cite{picozzi2007towards,picozzi2014optical}.

The DS stability is a necessary but insufficient condition for its existence because it develops from a stochastic process. The existing theories of the passive mode-locking self-start predict many possible sources of DS formation, including mode beatings and hole burning, dynamic gain saturation, parasitic reflections, spectrally selected absorption, and continuous-wave instability \cite{risken1968instability,goodberlet1990starting,krausz1992femtosecond,komarov2002phase,soto2002continuous,kapitula1998stability}. The thermodynamic theory of the ML self-start describes the soliton appearance as a first-order phase transition, which is affected by the laser noises distributed over a resonator period \cite{gordon2002phase,gordon2006self}\footnote{A laser operates far from equilibrium therefore the analogy to thermodynamics is not literal \cite{haken1975cooperative}.}. The experimental results of a DS self-emergence are ambiguous: on the one hand, the chirp intimidates the ML due to the negative feedback caused by passive frequency modulation, while on the other hand, relaxation oscillations enhance the ML self-start \cite{pronin2012high,li2010starting}.

This paper explores applying thermodynamic theory to analyze DS within a CPO framework. Utilizing the complex cubic-quintic nonlinear Ginzburg-Landau equation \cite{van1992fronts,moores1993ginzburg}, we investigate the theoretical limits and practical thresholds of DS energy scalability. Experimental approbation is carried out using a Cr$^{2+}$:ZnS CPO \cite{Rudenkov}, focusing on pulse energy scaling and the behavior under different parametric conditions. We aim to identify the principal limitations to energy scaling and define the conditions for maximizing DS energy in such systems. Our results illustrate that DS thermodynamic properties, such as entropy and temperature, influence the DS energy scalability. We identified the critical thresholds for DS energy scaling: DS entropy increase and transit to a non-equilibrium phase characterized by the negative temperature that leads to a DS splitting (multiple DS ``thermalization'').

\section{DS thermodynamics: adiabatic theory}\label{th}

The Ginzburg-Landau equation is a fundamental framework for exploring the nonlinear dynamics of various physical systems, including optics, condensed matter physics, nonlinear oceanic waves, Bose-Einstein condensates (BEC), and others  \cite{scott2006encyclopedia}. The most comprehensive formulations of this equation incorporate dissipative elements and have soliton-like solutions known as DS \cite{ankiewicz2008dissipative}. The latter describes the ultrashort pulses produced by the ML oscillators \cite{grelu2012dissipative} and is particularly noteworthy. 

Our subsequent analysis will utilize the following version of the complex cubic-quintic nonlinear Ginzburg-Landau equation \cite{podivilov2005heavily}:
\begin{gather}
 \frac{\partial}{\partial z} a\! \left(z,t\right)=-\sigma a\! \left(z,t\right)+\left(\alpha+i \beta\right) \frac{\partial^{2}}{\partial t^{2}} a\! \left(z,t\right)-\nonumber \\
 -i \gamma P\! \left(z,t\right) a\! \left(z,t\right)+\kappa\left(1-\zeta P\! \left(z,t\right)\right) P\! \left(z,t\right) a\! \left(z,t\right).   \label{eq:NGLE}
\end{gather}

\noindent Here $ a\! \left(z,t\right)$ is a laser field amplitude slowly varying with a local time $t$\footnote{Eq. (\ref{eq:NGLE}) is derived in a frame, moving with the group velocity of the pulse central (``carrier'') frequency. Likewise, the frequency $\omega$ in the subsequent analysis is a detuning from that central frequency.}, $z$ is a laser oscillator round-trip number, $\sigma$ is a saturated net-loss coefficient, $\alpha$, and $\beta$ are the squared inverse spectral filter bandwidth and GDD coefficients, respectively. $\gamma$ and $\kappa$ are the self-phase (SPM) and self-amplitude (SAM) modulation coefficients, respectively. $\zeta$ is a coefficient characterizing the saturation of self-amplitude modulation. $P \left(z,t\right) =  |a\! \left(z,t\right)|^2$ is a slowly varying power.

Some physical properties of DS could be qualitatively outlined in the following way. The phase balance between the contribution of normal GDD and SPM defines the DS wavenumber $q = \beta \Delta^2 =\gamma P_0$, where $2\Delta$ and $P_0$ are the DS spectral width and peak power, respectively \cite{podivilov2005heavily,kalashnikov2024dissipative}. That is analogous to the Schr{\"o}dinger soliton area theorem $q = \gamma P_0/2 =|\beta| T^2/2$ with a soliton width $\approx2T$ \cite{renninger2010area}. In contrast to the Schr{\"o}dinger soliton, a resonance with dispersive linear waves obeying the wavenumber $k = \beta \omega^2$ (a Langmuir
dispersion relation \cite{robinson1997nonlinear}) is possible for normal GDD $\beta >0$, when $k = q$. The existence of such resonance requires the cut-off point of the DS spectrum at $\omega^2 = \Delta^2 = \gamma P_0/\beta$, a spectral condensation $k \to 0$ by analogy with weak turbulent waves, and the subsequent formation of a Lorentzian-shaped truncated spectrum \cite{robinson1997nonlinear}. The detailed analysis confirms the latter conjecture \cite{podivilov2005heavily,kharenko2011highly,kalashnikov2024dissipative}.

A sole resonance condition does not provide the stability of the DS, and the analysis demonstrates a ``leakage'' energy outflow $k \to \infty$, so that this DS is only quasi-stable without an additional stabilization mechanism \cite{kalashnikov2023thermodynamics}. Spectral filtering ($\alpha > 0$ in Eq. (\ref{eq:NGLE})) provides such stabilization, preventing both spectral and temporal spreading of DS \cite{haus1991structures,chong2008properties}. The resulting additional losses must be compensated for by SAM so that $\alpha \Delta^2 \sim \kappa P_0$. Combining this condition with the resonance one results in $C \equiv  \left(\alpha/\beta \times \gamma/\kappa \right) \sim 1$, or ``soliton condition'' \cite{katz2006non}\footnote{$C \approx 1/3$ in \cite{chang2008dissipative} for the notations of (\ref{eq:NGLE}).}. As will be shown below, the $C$-parameter (``control parameter''), relating mutually dissipative and nondissipative factors in a system, defines a DS parametric space. Moreover, within this condition, the large chirp %
$\psi \approx 3/(\alpha/\beta+\kappa/\gamma) \gg 1$
\cite{kalashnikov2005approaching} requires strong domination of the nondissipative factors over the dissipative ones, i.e., $\gamma \gg \kappa$ and $\beta \gg \alpha$ \cite{podivilov2005heavily}. Those are the basic assumptions for the adiabatic theory of DS.

The \textit{adiabatic theory} \cite{podivilov2005heavily,kharenko2011highly} describes DS under the following assumptions:\textbf{ i)} $\beta >0$ in (\ref{eq:NGLE}), i.e., GDD is ``normal''; \textbf{ii)} non-dissipative factors described by $\beta$ and $\gamma$ coefficients prevail over the dissipative ones ($\alpha$ and $\kappa$); \textbf{iii)} as a result of large chirp $\psi \gg 1$, the field envelope $\sqrt{P\! \left(t\right)}$ evolves slowly with $t$ in comparison with the instant phase  $\varphi\left(t\right)$.

From these assumptions, one may obtain the explicit expressions for the DS spectral power $p(\omega)$ and energy $E$ (see Appendix \ref{secA1}):
\begin{gather} 
    p\! \left(\omega\right)=\frac{6 \pi \gamma  }{\zeta \kappa \left(\Xi^{2}+\omega^{2}\right)}\mathcal{H}\! \left(\Delta^{2}-\omega^{2}\right), \label{eq:spectrum}\\ 
E=\frac{6 \gamma \tan^{-1}\! \left(\frac{\Delta}{\Xi}\right)}{\zeta \kappa \Xi}, \label{eq:energy}   
\end{gather}

\noindent where $\omega$ is the frequency deviation from the pulse central frequency and $\mathcal{H}$ is the Heaviside function. 
Eq. (\ref{eq:spectrum}) describes a Lorentzian spectral profile of DS with the Lorentzian width $\Xi$ (Eq. (\ref{eq:xi})) truncated at $\pm \Delta=\sqrt{\gamma P_0/ \beta}$. The DS spectral profile (\ref{eq:spectrum}) is a good approximation for the partial exact solution of (\ref{eq:NGLE}) \cite{PhysRevA.77.023814} in the limit of $\psi \gg 1$, but without the strict conditions on the system parameters. The $\Delta$-parameter defines a DS spectral half-width measurable in an experiment \cite{Rudenkov}.

\begin{figure}[ht]
\centering
\centering\includegraphics[width=0.35\textwidth]{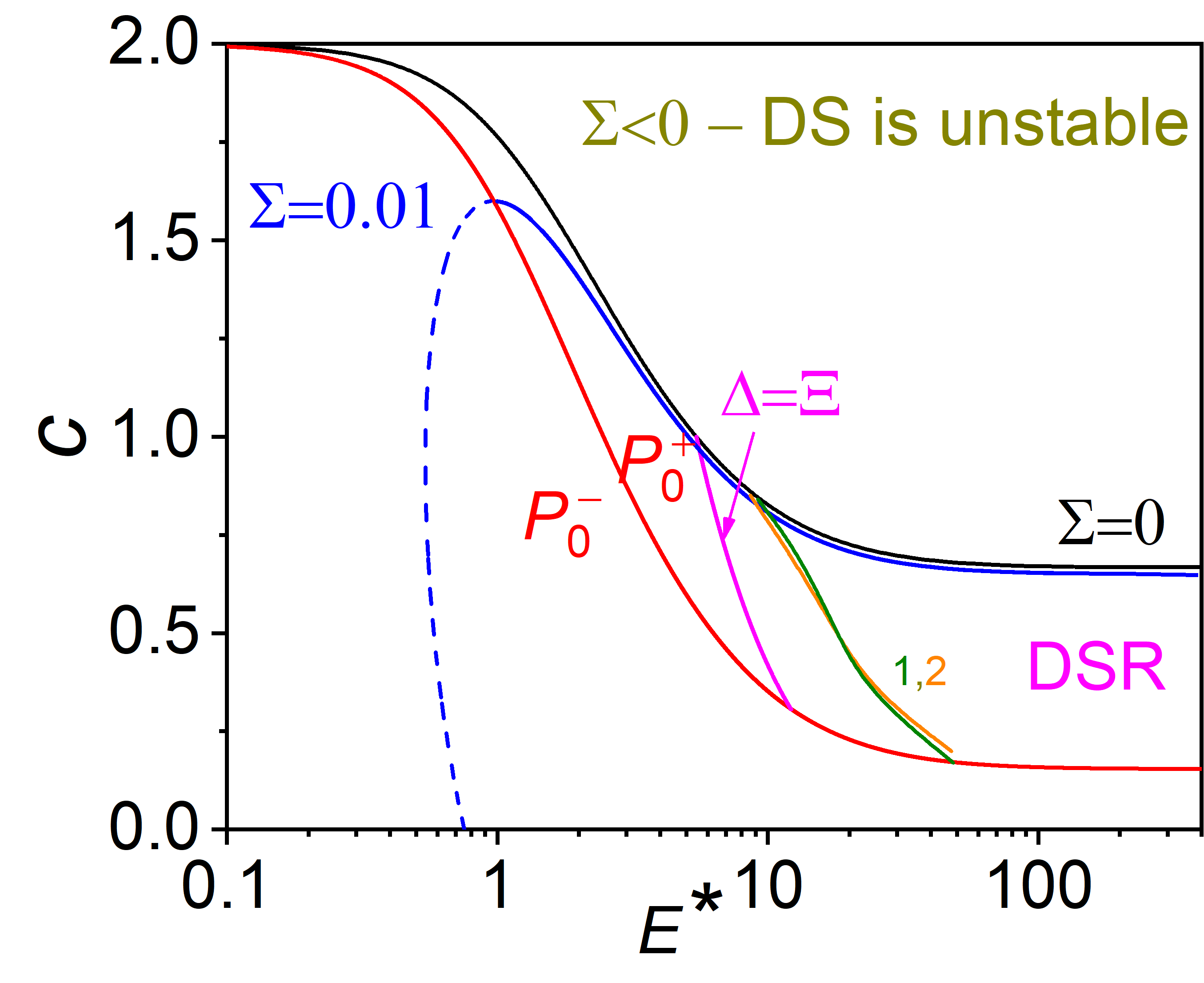}
\caption{DS master diagram. The black curve corresponds to a vacuum stability threshold $\Sigma=0$. DS is unstable above this curve. $\Sigma>0$ below this curve, and only one isogain $\Sigma=0.01$ for the positive $P_0^{+}$ (solid blue curve) and $P_0^{-}$ (dashed blue curve) is shown (see Eq. (\ref{eq:power})). The isogains shift along the red curve to the right with growing $E^\star$. The red curve divides the positive ($P_0^{+}$) and negative ($P_0^{-}$) branches of DS. The magenta curve (DS fidelity curve) $\Xi = \Delta$ marks a region of DSR. To the left of it, energy increase results in the DS power growth and its shortening ($\Delta$ increase). To the right of the magenta curve, higher energy is accommodated by DS broadening in time with $P_0$ and $\Delta$ tending to a constant. $\Xi$ tends to zero for the latter case, corresponding to a spectral condensation akin to BEC. (1,2) curves show the limit of energy scaling defined by DS splitting: (1): $\kappa=0.5$, $\zeta=1$ MW$^{-1}$ and (2): $\kappa=0.5$, $\zeta=2$ MW$^{-1}$; $\gamma=5.1$ MW$^{-1}$. $\alpha$ corresponds to 200 nm Cr:ZnS active medium gain bandwidth centered at 2.27 $\mu$m \cite{Rudenkov}. The energy $E$ is normalized: $E^{\star} =E \times \kappa \sqrt{\zeta/\beta\gamma}$. $C=\alpha \gamma/\beta \kappa$.}
\label{fig:fig1}
\end{figure}

Since the spectral parameters $\Delta$ and $\Xi$ can be expressed through the control parameter $C=\alpha \gamma/\beta \kappa$ and $\Sigma=4 \zeta \sigma/\kappa$ (Appendix \ref{secA1}), Eq. (\ref{eq:energy}) allows representing the DS parametric space in the form of the two-dimensional ``\textit{master diagram}'' (MD, Fig. \ref{fig:fig1}) \cite{kalashnikov2006chirped,Rudenkov,kalashnikov2024dissipative}. MD consists of a manifold of the ``isogains'' representing the $C$-parameter dependence on the dimensionless energy $E^{\star} =E \times \kappa \sqrt{\zeta/\beta\gamma}$ for $\Sigma=const$. $\Sigma$ can be connected with the average power $P_{av}$ in the continuous-wave (CW) regime of a laser through the approximation 
\begin{equation}
    \sigma \approx \vartheta (E/E_{cw}-1), \label{sig}
\end{equation}

\noindent which is valid in the vicinity of the lasing threshold \cite{kalashnikov2006chirped}. Here, $E_{cw} = P_{av} T_{cav}$ is a CW energy ($T_{cav}$ is an oscillator period) and $\vartheta=L^2/g_0$ ($L$ is a net-loss coefficient and $g_0$ is a small-signal gain). Threshold $\Sigma=0$ (black curve in Fig. \ref{fig:fig1}) divides the regions of DS stability against a CW-generation ($\Sigma>0$) and a CW-generation (``vacuum instability'' region, $\Sigma<0$). In practice, the multiple pulse generation could appear for the slightly positive $\Sigma$. The region of a vacuum-stable DS is divided into the sub-regions (Fig. \ref{fig:fig1}, red curve) corresponding to two different DS solutions of Eq. (\ref{eq:NGLE}): $P_0^{+}$ and $P_0^{-}$ (see Appendix \ref{secA1}, Eq. (\ref{eq:power})), which differ by their energy scalability. One isogain curve (blue) in Fig. \ref{fig:fig1} shows an example of this difference. The solid and dashed ones correspond to $P_0^{+}$- and $P_0^{-}$-branches, respectively. One can see that the limited energy scaling for the latter can be achieved only by reducing $C$ by, e.g., increasing the GDD.

The energy scalability of DS above the red curve in Fig. \ref{fig:fig1} illustrates the \textit{dissipative soliton resonance} (DSR) \cite{PhysRevA.78.023830,kalashnikov2024dissipative} \footnote{We define DSR as $\exists C^\star: \lim_{C \to C^\star} E = \infty$, i.e., there exists a set of $C$-parameters providing infinite energy asymptotic. The DSR phenomenon mimics a weakly dissipative BEC when $\Xi \to 0$ means a spectral condensation \cite{kalashnikov2021metaphorical}. Energy scaling due to the DS width growth is analogous to the BEC mass scaling due to the transition of a ``basin'' to the BEC phase. One has to note that the GDD term in (\ref{eq:NGLE}) corresponds to a boson kinetic energy if the SPM term describes a repelling two-boson interaction. We will discuss below the limitations of the analogy with the BEC.\label{fnDSR}}. 

The next characterizing curve belonging to the MD is a ``fidelity curve'' $\Xi=\Delta$ (magenta curve in Fig. \ref{fig:fig1}), which corresponds to a DS with a smoothed frequency dependence of the chirp ($\sim 1/(1-\omega^4/\Delta^4)$, Eq. (\ref{eq:sa}))   \cite{podivilov2005heavily}, i.e., a nearly constant chirp in the time domain. This enables DS compressibility outside of CPO by an anomalous GDD $\approx -\frac{3 \gamma}{2\kappa \beta} \Delta^{-4}$ \cite{kalashnikov2024dissipative} down to $\sim 1/\Delta$ with a minimal energy loss (see Subsection \ref{fidelity}) \cite{zhu2013generation}.

The fidelity curve characterizes a bottom energy threshold of DSR. The crossing of this threshold manifests itself by three experimentally recognizable marks \cite{Rudenkov}: \textbf{i)} the $T$-decrease with energy changes by increase; \textit{ii)} the DS spectral width $\Delta$ tends to a constant; and \textbf{iii)} the Lorentzian-shaped ``finger'' defined by the $\Xi$-decrease appears at a spectrum center. The blue curve in Fig. \ref{fig:fig2} with the corresponding Lorentzian fitting for a ``finger'' shows the referenced experimental spectrum.

Expression (\ref{eq:spectrum}) is a truncated Lorentzian spectrum with the width $\Xi=\sqrt{\gamma\big[(1+C)/\zeta-5P_0/3\big]/\beta}$. Let us consider the corresponding autocorrelation function, which resembles a damped oscillator\footnote{The relevant expression for the autocorrelation function of a strongly chirped DS solution $\sqrt{P_0}\sech(t/T)^{(1+i\psi)}$ of (\ref{eq:NGLE}) with $\zeta=0$ is presented in Appendix \ref{secA2}.}:
\begin{equation}
R(\tau) = \frac{3 \gamma \Delta}{2 \Xi \zeta \kappa} \int_{-\infty}^{\infty} e^{- \Xi | t |} \, \sinc\left( \Delta (\tau - t) \right) \, dt. \label{eq:autoc}
\end{equation}
\noindent Eq. (\ref{eq:autoc}) defines two characteristic time scales: the ``internal graining'' time $l=\pi/\Delta$ and the ``confining potential'' time $\Lambda=1/\Xi$. The first characterizes the DS spectral width tending to $\sqrt{\gamma/\beta \zeta}$, and the second defines the DS temporal width, which tends to infinity for DSR with the accompanying spectral condensation $p(\omega) \to \omega^{-2}$ ($\Xi \ll \Delta$) in the vicinity of $\omega=0$, and $R(\tau) \to \frac{3\pi \gamma \Delta}{2 \Xi \zeta \kappa} e^{- \Xi | \tau |}$.

Further, it is helpful to use the dimensionless values of Fig. \ref{fig:fig1}, namely, $E^*=E\times \kappa \sqrt{\zeta/\gamma\beta}$, $\omega^*=\omega \times \sqrt{\zeta \beta/\gamma}$, and $p^*(\omega^*)=p(\omega)\times \kappa/\beta$. Then, the spectral power looks as 
\begin{gather}\label{eq:spectrumn}
     p^*\! \left(\omega^*\right)=\frac{6 \pi }{ \left(\Xi^{*2}+\omega^{*2}\right)}\mathcal{H}\! \left(\Delta^{*2}-\omega^{*2}\right),\nonumber \\
     \Xi^{*2}=1+C-\frac{5}{3}\Delta^{*2}\\ \nonumber
\Delta^{*2}=\frac{3}{4}\left[ 1 - \frac{C}{2} \pm \sqrt{\left( 1-\frac{C}{2} \right)^2-\Sigma}\right],
\end{gather}

\noindent which is akin to the Rayleigh-Jeans spectrum (momentum) distribution of a semi-incoherent soliton \cite{picozzi2014optical} and turbulence \cite{nazarenko2011wave}. Indeed, such a spectrum results from the high chirp $\psi$ (see the proposition \textbf{iii)} above), leading to the DS phase inhomogeneity in our case. The interpretation of the chirped DS as a semi-incoherent soliton being a microcanonical statistical ensemble of ``quasi-particles'' defined by the ``graining time'' $l$ and confined by potential with the characteristic time scale $\Lambda$ in Eq. (\ref{eq:autoc}) \cite{akhmediev2000multi,picozzi2007towards,picozzi2009thermalization,picozzi2014optical}, allows using the ideology of statistical mechanics to the DS phenomenology \cite{wu2019thermodynamic,kalashnikov2024dissipative}. When DS transits to the energy scaling regime closely interrelated with the spectral condensation $p(\omega) \to \omega^{-2}$ ($\Xi \ll \Delta$), the statistical approach becomes reasonable. The tendency to the decoupling of ``quasi-particles'' is enhanced due to the decoupling of the characteristic time scales $l \ll \Lambda$ when $R(\tau) \to \frac{3\pi \gamma \Delta}{2 \Xi \zeta \kappa} e^{- \Xi | \tau |}$ (see (\ref{eq:autoc})).

Let us consider (\ref{eq:spectrum}) as a probability distribution in the momentum space \cite{wu2019thermodynamic,kalashnikov2024dissipative} and assume a \emph{soliton condition} (``potential condition'') $C \simeq 1$\footnote{The limiting value of $C$ for DSR is $C=2/3$ when $\Sigma=0$. It should be noted that $\beta-$sign is opposite to that in the nonlinear Schr\"{o}dinger equation. Therefore, this condition is soliton-\emph{like}.} implying a Gibbs-like steady-state probability distribution in statistical mechanics \cite{katz2006non}. Following this intention, one could define the DS ``chemical potential'' $\mu$, ``entropy'' $H_s$, ``internal energy'' $U$, and ``temperature'' $\Theta$ (see Appendix \ref{secA3}) \cite{wu2019thermodynamic,kalashnikov2024dissipative}: 
\begin{gather} 
   -\mu=\Xi^2,\\
   H_s=\dfrac{1}{\tan^{-1}\left( \dfrac{\Delta}{\Xi} \right)} \int_{0}^{\Delta/\Xi} \dfrac{\ln(1 + x^2)}{1 + x^2} \, dx,\\
   U =\dfrac{\Xi}{2 \tan^{-1}\left( \dfrac{\Delta}{\Xi} \right)} \int_{-\Delta}^{\Delta} \dfrac{\omega^2}{\omega^2 + \Xi^2} \, d\omega,\\
   \Theta = \left( \frac{\partial H}{\partial U} \right)^{-1}.\label{eq:entr}
\end{gather}

\section{Results}\label{res}

\begin{figure}
    \centering
    \includegraphics[width=0.9\linewidth]{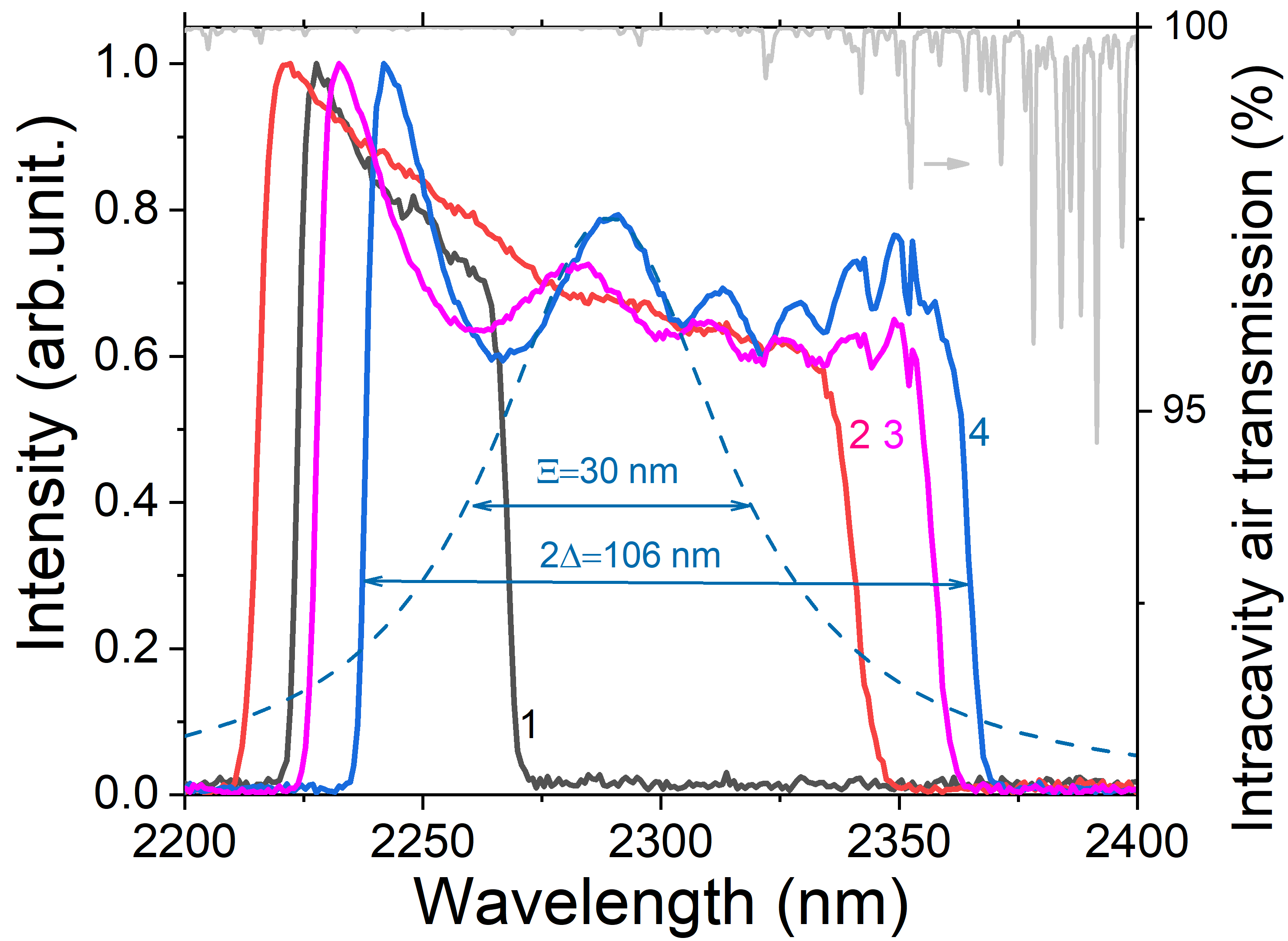}
    \caption{Experimental spectra (solid lines, left scale) from a Cr$^{2+}$:ZnS CPO \cite{Rudenkov}. The output coupler transmission, average power, and repetition rate are 14\%, 272 mW, and 12.3 MHz, respectively. The output DS energies are 5.1 (black, 1), 16.1 (red, 2), 19.2 (magenta, 3), and 22 (blue, 4) nJ. The dashed blue curve is a Lorentzian fitting of the experimental spectrum's central spike (``finger''). The spectrum asymmetry and shift can be attributed to the contribution of the third-order GDD in combination with the asymmetrical spectral dissipation \cite{sorokin2023atmospheric}. The gray line (right scale) shows the intracavity atmospheric transmission over the full round-trip, calculated with the same resolution as the experimental spectra.}
    \label{fig:fig2}
\end{figure}

We define the DSR region as an area on MD (Fig. \ref{fig:fig1}) confined by the conditions $\Xi < \Delta$, that means $\Sigma<\frac{3}{16}\left( 1-C^2 \right)$, and $C \in \left[ 2-4\sqrt{\Sigma},2/3 \right] $. The energy scaling within this region does not require more GDD because the $C$-parameter could be constant. A decrease of $\Xi$ compared to $\Delta$ means a transition to the spectral condensation regime, in which the DS harvests the energy by becoming longer with the simultaneous tending of $\Delta$ and $P_0$ to constant values. The signatures of transit to the condensation phase are visible in the experiment. Fig. \ref{fig:fig2} demonstrates the modification of the DS spectral profile with energy growth. One can see that spectral broadening saturates with energy. At the same time, a Lorentzian central spectral spike (blue curve) appears, as indicated by the dashed blue curve fit. The spike is accompanied by a low-frequency modulation of the spectrum, which is visible on its long-wavelength wing (spectra 2 and 3 in Fig. \ref{fig:fig2}). Such distortions are typical for the broadband CPO \cite{kalashnikov2006chirped,kalashnikov2007spectral}. We associate them with an excitation of the ensembles of ``quasi-particles''\footnote{Such distortions can be described in the framework of the perturbation theory in spectral domain developed in \cite{kalashnikov2011dissipative} (see Appendix \ref{secAperturb}).}, which will eventually destabilize the DS if the energy increases further. Note that the sharp modulation, visible at the long-wavelength side of the spectra, is due to the water vapor absorption in the atmosphere inside the cavity. The form and amplitude of this modulation are well described by the analytical theory \cite{CPOhair}, which predicts enhancement of the amplitude near the edges of the DS spectrum, as seen in Fig. \ref{fig:fig2}. This modulation does not influence DS stability or other properties.

\subsection{DS fidelity and compressibility}\label{fidelity}

The energy scaling provided by the spectral condensation, i.e., the $\Xi$-decrease, decouples the correlation scales $l$ and $\Lambda$. The first consequence of a shift from the fidelity curve with an asymptotic energy growth causing $l/\Lambda \to 0$ is a spectral chirp inhomogeneity (see Fig. \ref{fig:fig3} and Eq. (\ref{eq:sa}) in Appendix \ref{secA1}). The spectral chirp, almost constant within the broadest spectral diapason for $\Xi=\Delta$, becomes strongly distorted with further energy growth when $\Xi/\Delta$ decreases.  

\begin{figure}
    \centering
    \includegraphics[width=0.85\linewidth]{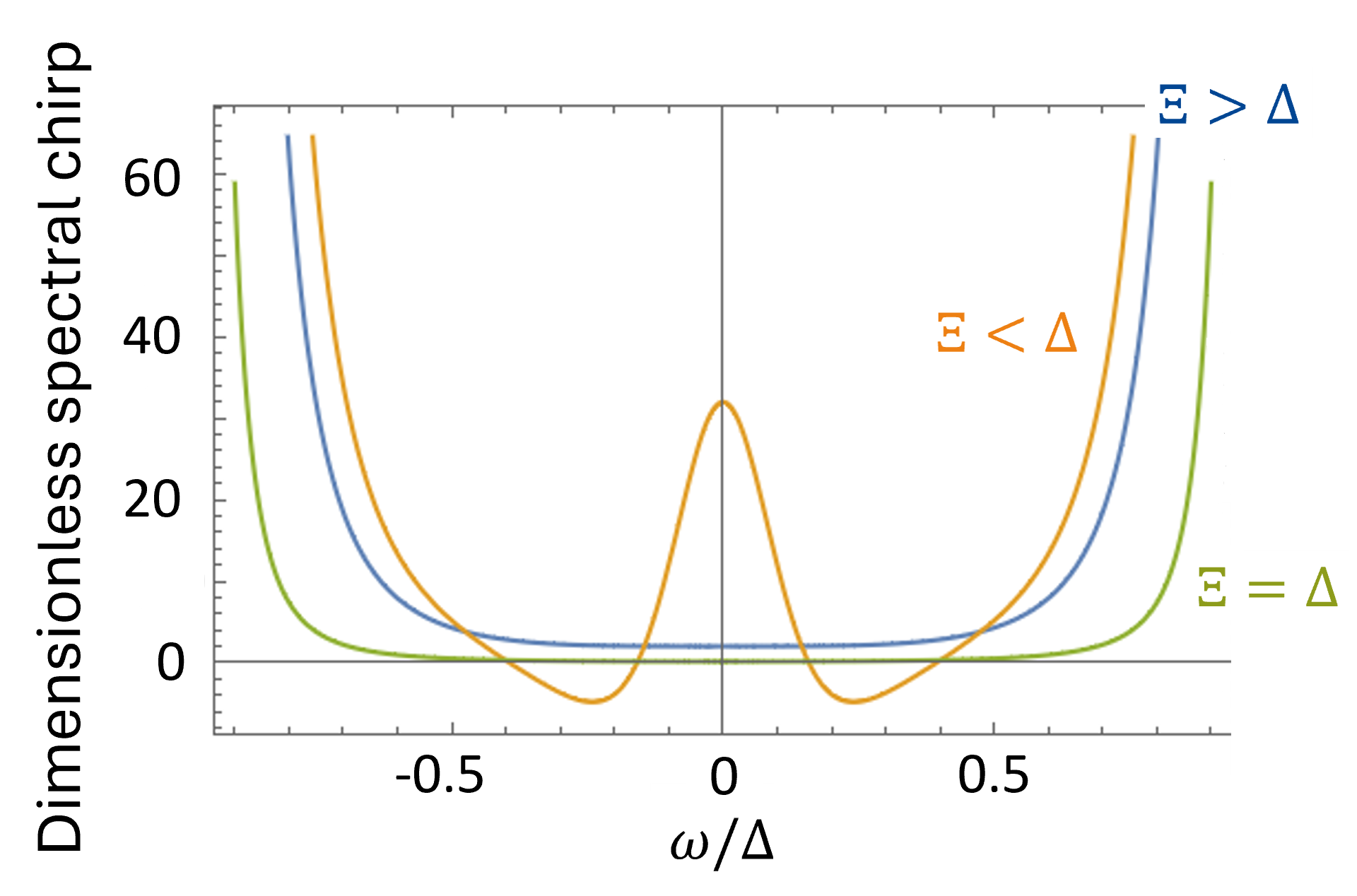}
    \caption{Dimensionless spectral chirp $\psi(\omega)\propto\frac{\partial^2 }{\partial \omega^2}\frac{\omega^2}{\left( \Xi^2+\omega^2 \right)\left( \Delta^2-\omega^2 \right)}$ vs. $\omega$ expressed in the units of $\Delta$ for three different relations between $\Xi$ and $\Delta$.}
    \label{fig:fig3}
\end{figure}

\begin{figure}
    \centering
    \includegraphics[width=1\linewidth]{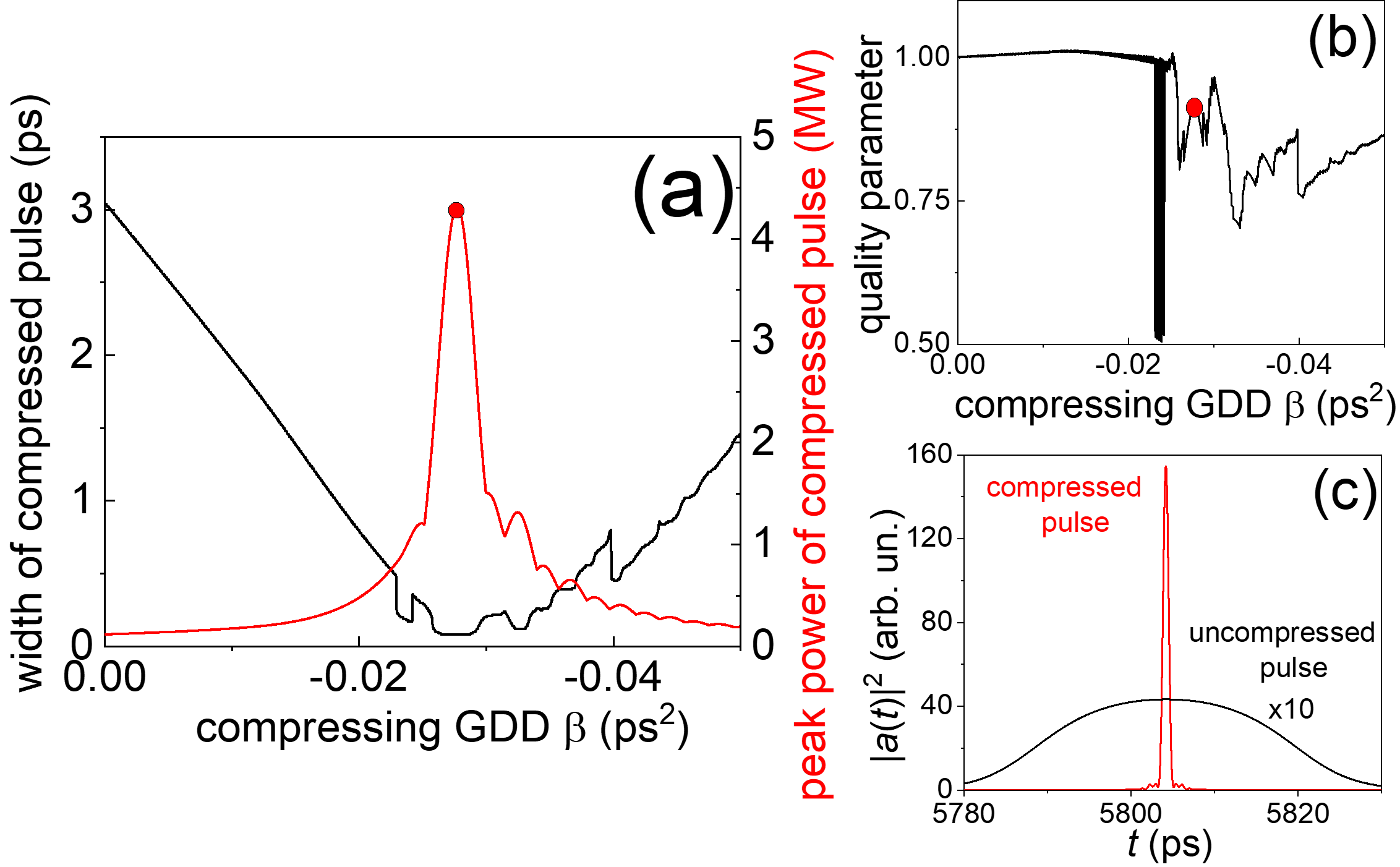}
    \caption{(a): Dependencies of the FHWM width (black) and peak power (red) of a compressed DS on the anomalous GDD $\beta$ of the compressor. (b): The corresponding dependence of the compression quality factor. (c): DS profile after CPO (black, ten-fold scaling) and after maximum compression (red), corresponding to the red point on graph (b). $C=0.72$, $E^*=11.6$, $\kappa=0.5$ MW$^{-1}$, $\zeta=1$ MW$^{-1}$, other parameters same as  for Figs. \ref{fig:fig1},\ref{fig:fig2}.}
    \label{fig:fig4}
\end{figure}

The calculated results of the chirped DS compression by varying the frequency-independent GDD are shown in Fig. \ref{fig:fig4}. The quality parameter is a ratio of the energy confined within the FWHM (full width at half maximum) region of the input DS to that of the compressed pulse \cite{travers2011ultrafast}. The quality parameter remains close to unity, and the compressed pulse width decreases monotonously with anomalous dispersion till the value of the latter approaches $-3 \gamma^2/2\beta \kappa \zeta \Delta^4 \approx -0.02$ ps$^2$  \cite{podivilov2005heavily,kalashnikov2024dissipative}. Further increase of anomalous dispersion causes pulse fragmentation with the inevitable degradation of compression quality. However, the minimal pulse width of $\approx 80$ fs for $\beta \approx$-0.028 ps$^2$ can be achieved for the quality parameter $\approx$0.9 at a cost of oscillating pulse wings (Fig. \ref{fig:fig4}).

\subsection{DS statistics}\label{stat}

To investigate the statistics of DS emerging from quantum noise, we performed numerical simulations according to Eq. (\ref{eq:NGLE}) supplemented by an additive complex Gaussian white noise term $\Gamma(z,m \Delta t)$ ($m \in  [1...N]$ is a cell number, $N=2^{16}$ with a 10 fs time-cell size) \cite{206583,kalashnikov2023thermodynamics}:
\begin{gather} \label{eq:noise}
\left\langle \Gamma_m (z)\Gamma_n^* (z') \right\rangle = W \delta_{mn}\delta(z-z') \nonumber \\ 
\left\langle \Gamma_m (z)\Gamma_n (z') \right\rangle = 0,\\
W = 2 h \nu |\sigma|/T_{cav}.\nonumber 
\end{gather} 

\noindent Here $W$ is a noise power and $\nu$ is a carrier frequency. $W$ is associated with a ``vacuum temperature'', and $T_{cav}$ is a cavity period. We assume the DS energy scaling by $T_{cav}$ so that a continuous-wave power $P_{av}$ is constant. The scaling parameter is a ``continuous-wave energy'' $E_{cw} = P_{av} T_{cav}$ (an ``isothermic'' process in the sense of \cite{gordon2006self}, see Eq.  (\ref{sig})).

Since multiple DSs can occur due to the excitation of both quantum noise and internal DS perturbations, it is convenient to consider the $E_{cw}^*$-variable instead of $E^*$. As shown in \cite{gordon2006self}, $E_{cw}^*$ plays a role in the environment ``temperature'' when a passive ML self-start is considered as the first-order phase transition caused by a ``vacuum heating'' or ``stochastic resonance'' \cite{RevModPhys.70.223}.

\begin{figure*}[htb]
  \centering
  \includegraphics[width=0.8\textwidth]{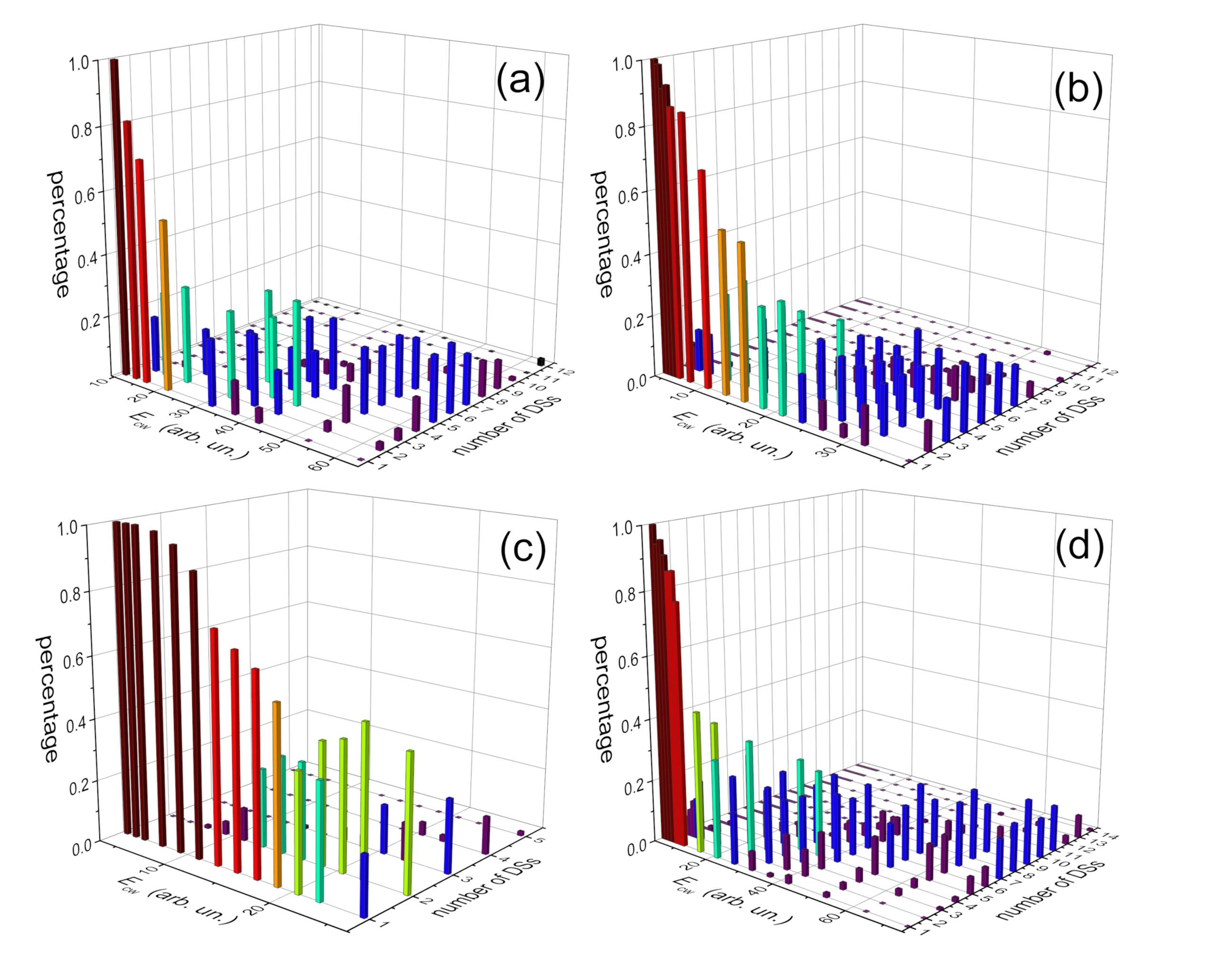} 
  \caption{Percentage of multiple DSs gathered from 150 independent stochastic samples. (a): $\kappa=\zeta=$1 MW$^{-1}$, $\vartheta=$0.04; (b): $\kappa=$1 MW$^{-1}$, $\zeta=$0.5 MW$^{-1}$, $\vartheta=$0.04; (c): $\kappa=$0.5 MW$^{-1}$, $\zeta=$1 MW$^{-1}$, $\vartheta=$0.04; (d): $\kappa=$1 MW$^{-1}$, $\zeta=$1 MW$^{-1}$, $\vartheta=$0.02. For all graphs $\gamma$=5.1 MW$^{-1}$, and $C=$0.27 which is kept constant by appropriate scaling of $\beta$.}
  \label{fig:fig5probability}
\end{figure*}

The examples of the multiple DS probability distribution harvested on 150 stochastic samples are shown in Fig. \ref{fig:fig5probability}. One may see a tendency for multipulsing with higher energy, leading to a ``thermalization'' of the DS, i.e., smoothing of the distribution function, which shifts to a higher number of DSs. Such thermalization limits the DS energy scaling. Curves 1 and 2 in Fig. \ref{fig:fig1} show the thresholds of the energy scaling, after which the probability of single DS generation from a quantum noise falls below 95\%.

\begin{figure}[htbp]
    \centering
    \begin{minipage}[b]{0.45\textwidth}
        \centering
        \includegraphics[width=\linewidth]{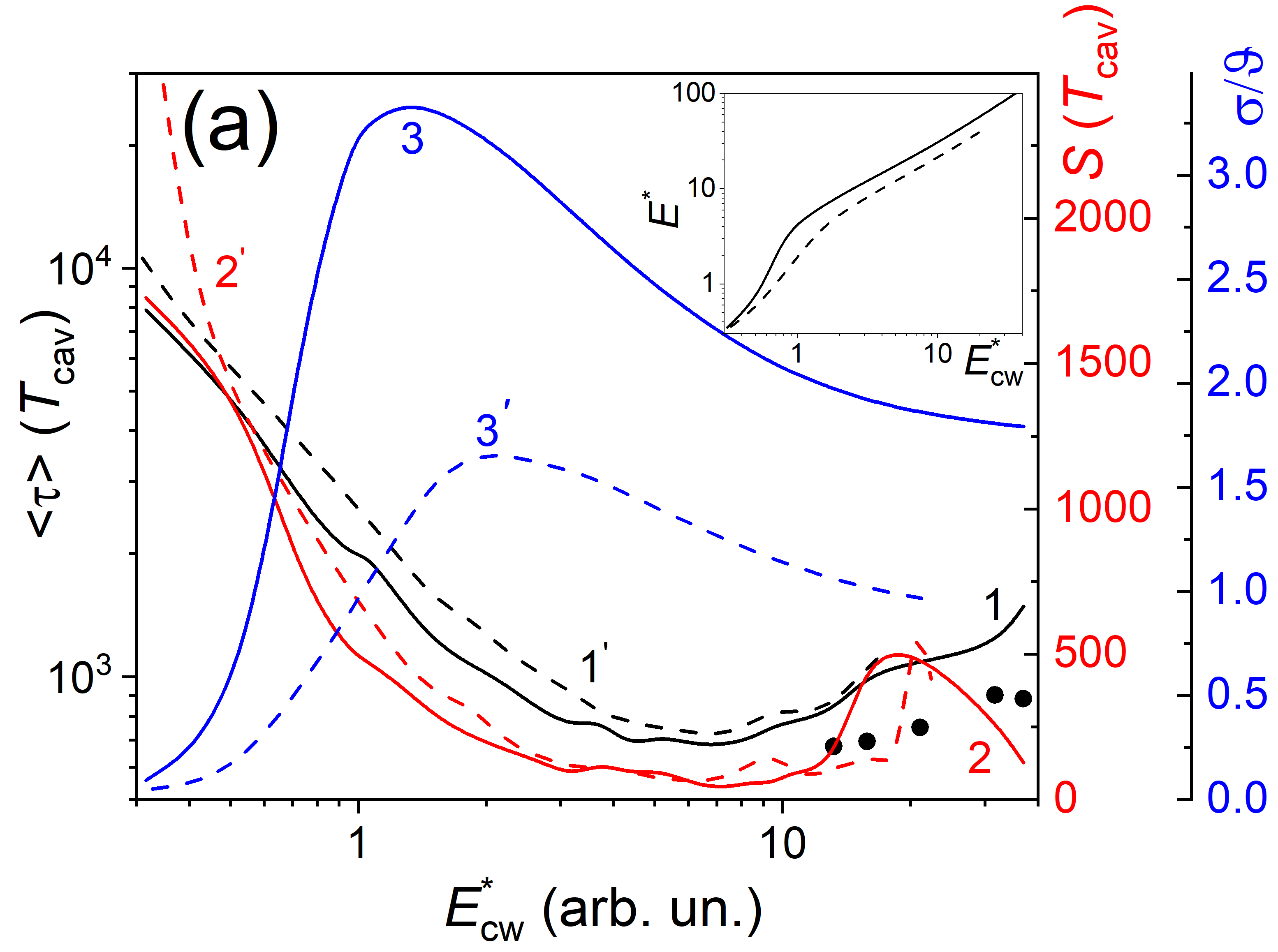} %
        \label{fig:fig6a}
    \end{minipage}
    \hfill
    \centering
    \begin{minipage}[b]{0.4\textwidth}
        \centering
        \includegraphics[width=\linewidth]{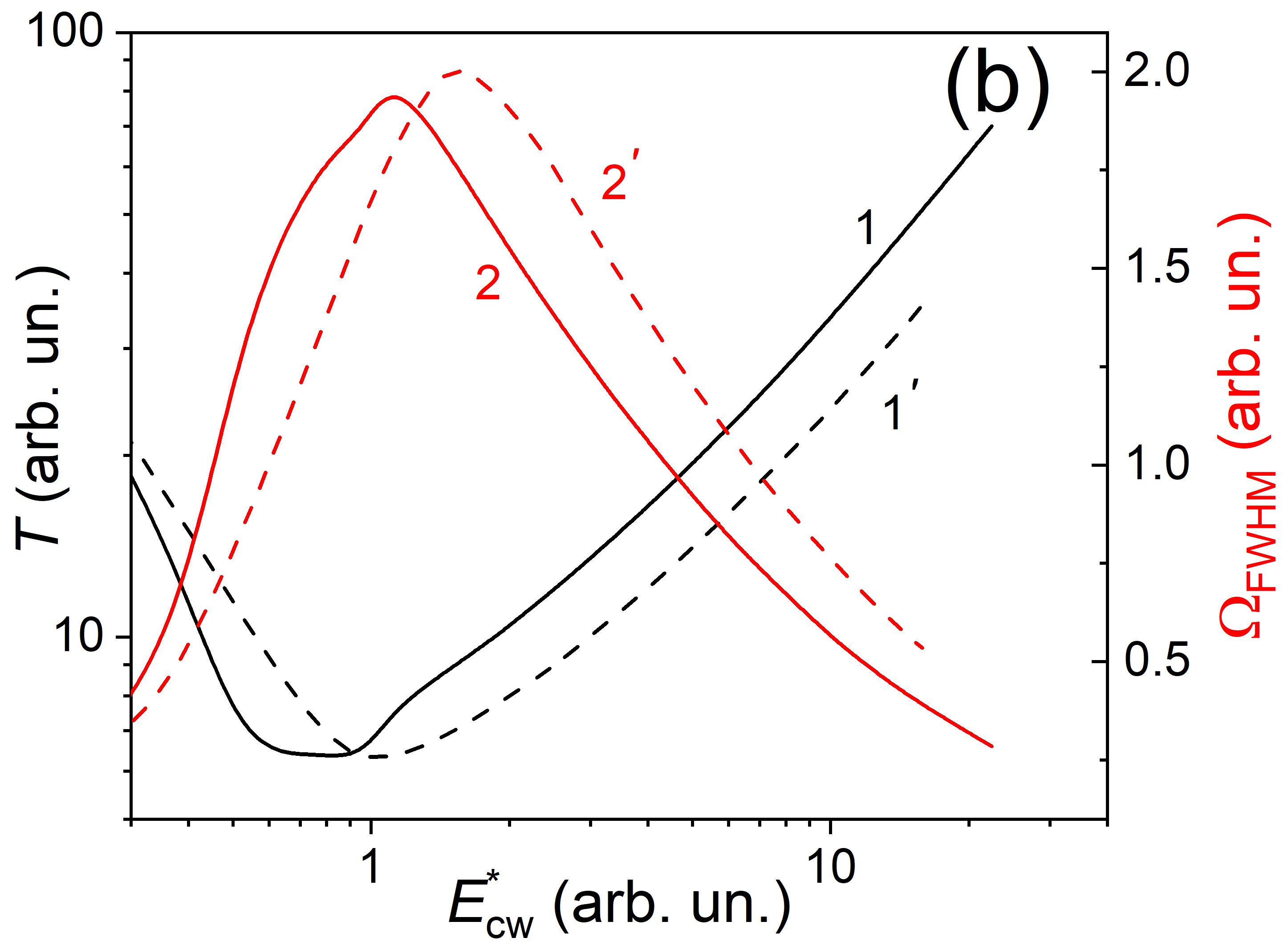} %
        \label{fig:fig6b}
    \end{minipage}
\caption{CW energy dependencies of (a): averaged DS self-start time $<\tau>$ (black curves 1), its standard deviations $S$ (red curves 2) and order parameter $\sigma/\vartheta$ (blue curves 3); (b): DS FWHM width $T$ (black curves) and FWHM spectral width $\Omega_\mathrm{FWHM}$ (red curves). Energy is normalized to $\kappa \sqrt{\zeta/\beta \gamma}$, frequency is normalized to $\sqrt{\beta \zeta/\gamma}$, and time is normalized to $\kappa/\sqrt{\beta \zeta \gamma}$. The physical parameters are close to those in Ref. \cite{Rudenkov}: (a) $\kappa=$ 1 MW$^{-1}$ (solid curves 1, 2, 3) and (b) 0.5 MW$^{-1}$ (dashed curves $1'$, $2'$, $3'$). $C$=0.54, $\gamma$=5.1 MW$^{-1}$,  $\zeta=$ 1 MW$^{-1}$, and spectral filter bandwidth equals 200 nm. Points show $<\tau>$ for the dual DSs, $\kappa=$ 1 MW$^{-1}$. The inset in (a) shows the connection between $E_{cw}^*$ and $E^*$ in Fig. \ref{fig:fig1} through the order parameter $\sigma/\vartheta$. The solid and dashed lines correspond to  $\kappa=$ 1 and 0.5 MW$^{-1}$, respectively.}
\label{fig:fig6}
\end{figure}

The simulated statistical characteristics of DS, i.e., (a): the averaged DS self-start time $\tau$ (black curves 1), its
standard deviation $S$ (red curves 2), order parameter $\sigma/\vartheta$ (blue lines 3), and (b): the DS FWHM width $T$ (black lines), and the FWHM spectral width $\Omega_\mathrm{FWHM}$ (red lines), are shown in Fig. \ref{fig:fig6}. The order parameter $\sigma/\vartheta$ \cite{Fischer:13} shows a level of the DS energy domination caused by SAM over the continuous wave. This parameter allows rescaling MD of Fig. \ref{fig:fig1} represented in ($E^*$ vs. $C$)-coordinates into the analogous diagram in ($E_{cw}^*$ vs. $C$)-coordinates by the transformation $E_{cw}^*=E^*/(1+\sigma/\vartheta)$ (see inset in Fig. \ref{fig:fig6},a). The experimental measurement of $E$, $E_{cw}^*$, and an approximate calculation of the SAM parameters involving the spectral and temporal DS widths may connect the experimental data with an MD representation \cite{Rudenkov}. $E_{cw}^*$ plays a role in the noise temperature in the thermodynamic theory of mode-locking \cite{gordon2006self}\footnote{It is directly connected with the quantum noise power $W$ in Eq. (11).}.

Fig. \ref{fig:fig6} demonstrates the following tendencies. i) The average self-start time (black curves) and its standard deviation (red curves) decrease initially with increasing $E_{cw}^*$, reach the minimum, and then increase. A two-pulse generation must appear starting from some energy, and its average self-start time (black points) is lower than that of a single-pulse DS. This corresponds to higher multipulsing probability at higher energies, following Fig. \ref{fig:fig5probability}. In parallel with the energy growth, the order parameter reaches a maximum and then decays (blue curves in Fig. \ref{fig:fig6}).

In parallel with these statistical signatures, the energy scaling by a DS power changes with that by a DS width $T$ (black curves in Fig. \ref{fig:fig6},b), which can be treated as a transition to the DSR regime \cite{Rudenkov}. In parallel to the order parameter maximization, the DS FWHM spectral width reaches a maximum and then decreases (red curves in Fig. \ref{fig:fig6},b). Such a maximum means $\Xi=\Delta$, i.e., maximum fidelity, with a subsequent $\Xi$-diminishing.

\subsection{DS ``quantization'' and thermolization}\label{quant}

The MD (Fig. \ref{fig:fig1}) is, in fact, a manifold of isogains $\Sigma=const \ge 0$ belonging to two branches of DS: $P_0^{+}$ (``energy scalable'' branch) and $P_0^{-}$ (``energy unscalable'' branch) ($\pm$-signs in Eq. (\ref{eq:power})). This means that the $P_0^{+}$-solution belongs to the DSR region, i.e., it has an infinite-energy asymptotic for some fixed $C$ and $\Sigma$. There is no asymptotic solution for the $P_0^{-}$. 

\begin{figure}[htbp]
    \centering
    \begin{minipage}[b]{0.38\textwidth}
        \centering
        \includegraphics[width=\linewidth]{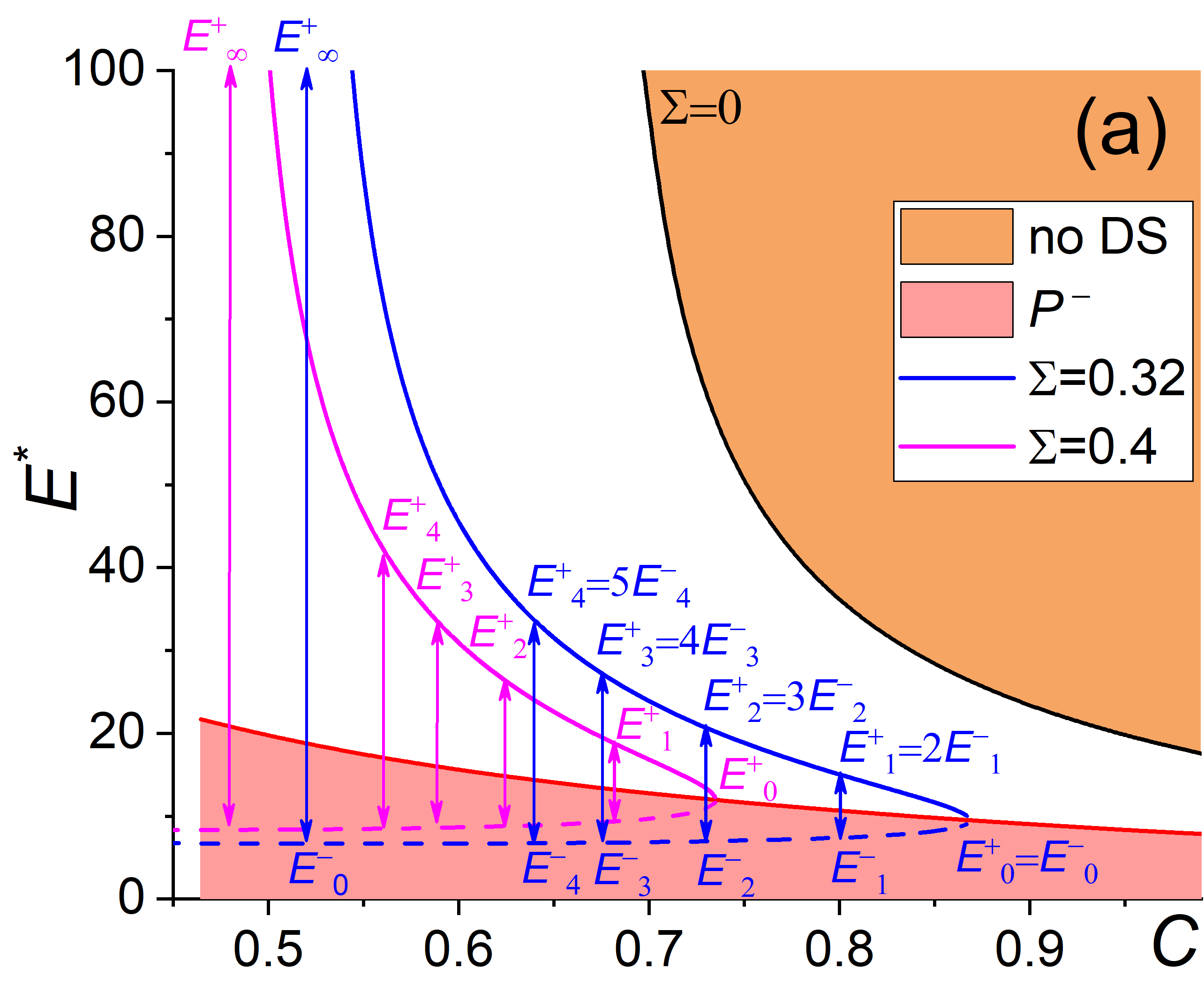} %
        \label{fig:fig7a}
    \end{minipage}
    \hfill
    \centering
    \begin{minipage}[b]{0.38\textwidth}
        \centering
        \includegraphics[width=\linewidth]{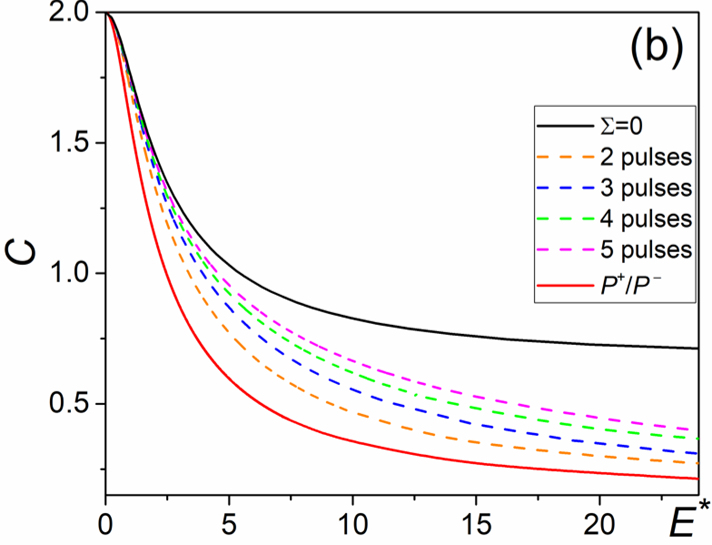} %
        \label{fig:fig7b}
    \end{minipage}
\caption{(a) Illustration of the DS energy pairing (see text). There is a pairing between $P_0^{+}$ DS and the corresponding $(n+1) P_0^{-}$-DS complex ($n=0,1,2,...$) with the same energy and $\Sigma$. The red curve is derived from Eq. (\ref{eq:power}) and indicates the boundary separating the two analytical DS branches corresponding to the energy scalable (DSR-capable) and energy unscalable solutions. (b) Borders of the DS existence (black solid curve) and the division between the $P_0^{+}$ and $P_0^{-}$ branches of DS (red); the dashed curves show the $C$ vs. $E^*$-dependencies of the $P_0^{+}$-branch DS having the counterpart complexes of two ($n=1$), three ($n=2$), four ($n=3$), and five (($n=4$)) $(n+1) P_0^{-}$-DSs with the same $\Sigma$ and total energy as their $P_0^{+}$-partner.}
\label{fig:q}
\end{figure}

The next physical meaning regarding the $P_0^{+}/P_0^{-}$-division will have an immediate consequence for DS stability. For fixed DS and CW energies, i.e., a fixed $\Sigma$, a ``bent'' behavior of the isogain curves (Fig. \ref{fig:fig1} and  Fig. \ref{fig:q},a) allows a $(n+1)$-set of identical $P_0^{-}$-pulses with the same total energy and $\Sigma$ as a single $P_0^{+}$-pulse (Fig. \ref{fig:q}, a). Here $n=0,1,...$ is a number of $P_0^{-}$ pulses conjugated with a $P_0^{+}$ DS ($P_0^{+}$ coincides with $P_0^{-}$ for $n=0$). In this sense, the $P_0^{+}$-DS allows energy ``quantization'' \cite{PhysRevA043816,vodonos2004formation,komarov2005multistability,renninger2010area,komarov2013competition} into a set of identical non-interacting $P_0^{-}$-DSs\footnote{The numerical calculations demonstrate that the pulses in a set are equal and well-separated in the vast majority of cases. However, the probability of the appearance of non-equal or interacting pulses increases with $\zeta$. Such dynamical regimes need further study \cite{grelu2024solitary}.}. For strongly chirped DS, the ``quantization'' is possible due to the existence of two DS domains:  $P_0^{+}$ providing DSR and $P_0^{-}$ without DSR (see Footnote \ref{fnDSR}). The vertical arrows in Fig. \ref{fig:q}, a show the interconnections between such states. 

Since the isogains with larger $\Sigma$ shift in the direction of higher $E^{*+}$ and lower $C$ (blue and magenta curves in Fig. \ref{fig:q}, a), it results in the well-defined $C-E^*$-dependencies for the $P_0^{+}\to (n+1)P_0^{-}$-correspondences (Fig. \ref{fig:q}, (b)), which are discrete in both $C$ and $E^*$ sections (orange--magenta curves in (Fig. \ref{fig:q}, (b)). The DSR with $E^{*} \to \infty$ could be treated as the $n \to \infty$ limiting level, where $n$ is the number of potentially excitable pulses. Then, the $P_0^{+}/P_0^{-}$-division curve corresponds to the ground state $n=0$ and the ``levels'' with higher $n$ are the ``excited'' ones.


However, the possibility of such a multipulse state does not imply its feasibility. Fig. \ref{fig:fig5probability} demonstrates the feasibility of such states, i.e., the transition to a multipulsing with the energy growth and, as a consequence of further energy growth, a thermodynamic thermalization, or equidistribution over high-order multipulse states.


%

Under ``thermalization,'' we understand the following. Defining $p(n)$ as the probability of $n$-pulses appearance, we observe in Fig. \ref{fig:fig5probability} the shift of its maximum towards higher $n$ values with increasing $E_{cw}^*$. However, the dissipative character of a system described by Eq. (\ref{eq:NGLE}) leads to a broadening of $p(n)$ with $E_{cw}^*$. $p(n)$ ``scatters'' over $n$ with the increasing $E_{cw}^*$, because the total energy $(n+1) E_n^{*-}$, where $E_n^{*-}$ is the energy of individual pulse in a set,
 is not necessarily equal to some $const$ for a given $E_{cw}^*$. This causes different sets of $(n+1) E_n^{*-}$ with different $n$ to be realizable for a fixed $E_{cw}^*$ if the latter is sufficiently large.

\subsection{DS entropy}\label{Hs}

\begin{figure}
    \centering
    \includegraphics[width=0.8\linewidth]{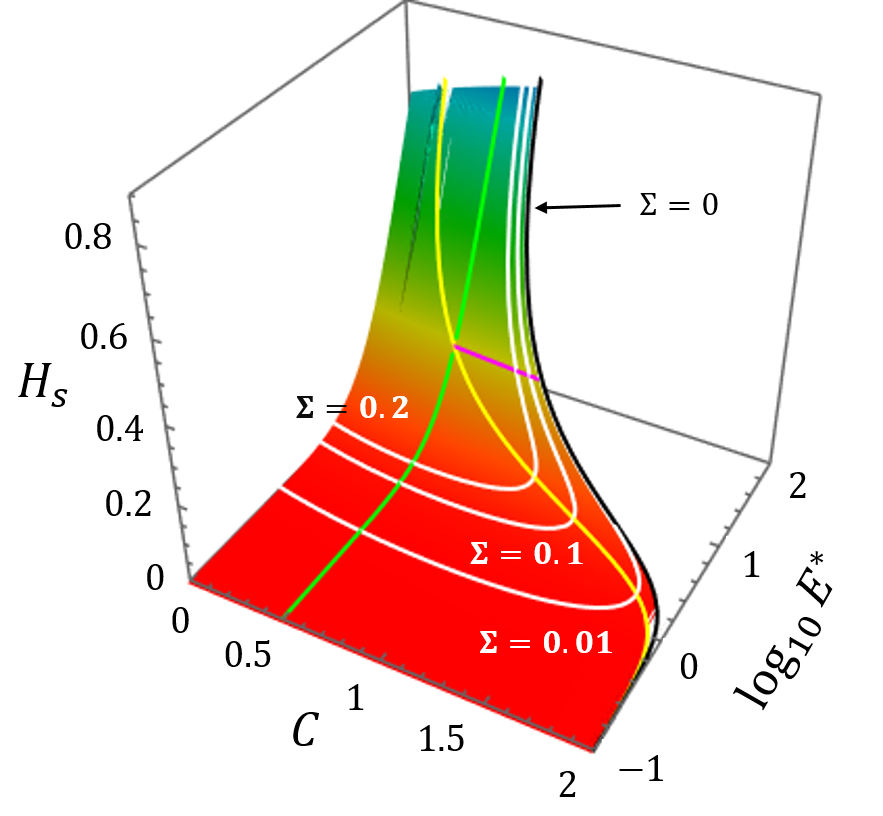}
    \caption{The DS entropy $H_s$ versus $C$ and the logarithm of $E^*$ for the DS solutions $P_0^{+}$ (right side of the surface) and $P_0^{-}$ (left side of the surface). The yellow curve shows the border between them. The black curve with $\Sigma=0$ confines a region of DS existence. The magenta line with the constant $H_s \approx 0.22$ represents a fidelity curve ($\Xi=\Delta$) in Fig. \ref{fig:fig1}. White curves show the isogains with the constant $\Sigma$. The green curve $C=0.5$ shows one of the curves with a constant $C$. The DSR region is located between the yellow, black, and magenta curves.}
    \label{fig:entropy}
\end{figure}

The DS energy growth results in $\Xi \to 0$ and $\Delta \to const$. In terms of the time correlation scales of Eq.(\ref{eq:autoc}), that means the multiplication of microstates and the growth of entropy. The entropy (8) of both DS branches, divided by the yellow curve, is shown in Fig. \ref{fig:entropy}. Curves demonstrate $H_s$-dependencies on $C$ and $E^*$ for fixed $\Sigma$ (white curves), $C$ (green curve), and the fidelity line (magenta). The entropy value for the latter is constant due to $\Xi/\Delta=1$ demonstrating a crucial role of the relation between an internal graining scale $\Lambda$ and a confining potential scale $l$ for the formation of microstates' ensemble (Eq. (\ref{eq:autoc}))\footnote{In this sense, all curves $\Xi/\Delta=const$ are iso-entropic.}.

Here, we must note a principal difference with BEC. Although DSR means a transit to the spectral condensation $\Xi \to 0$ when energy concentrates at a DS spectrum center $\omega=0$, the DS entropy rises, unlike BEC. For the latter, the occupation of a ground state means $H \to 0$. For DS, the spectral condensation means the growth of difference between two characteristic time scales $l$ and $\Lambda$, i.e., the growth of the microstate number and, consequently, the $H$-growth tending to $\ln{4}=1.39.$\footnote{In essence, such difference consists in the following. i) The sign of GDD in Eq. (\ref{eq:NGLE}) is opposite to that in the Gross-Pitaevskii equation, i.e., either a kinetic term corresponds to negative boson mass or a two-boson interaction is repelling. ii) Thus, DS formation is possible only due to dissipation, which results in the bounded spectrum $\omega^2 \le \Delta^2$ (Sec. \ref{th}). The latter causes an emergence of two correlation scales in (\ref{eq:autoc}). The difference between them increases with energy that increases entropy since we relate the microstates with the smallest scale $l$ \cite{jordan2000mean} so that their quantity could be qualitatively related to $\Lambda/l$.} Therefore, it is more adequate to make the analogy with the incoherent solitons \cite{
picozzi2014optical,picozzi2009thermalization,picozzi2007towards} and turbulence \cite{picozzi2014optical,nazarenko2011wave,robinson1997nonlinear}, in particular, the process of entropy growth when the formation of large-scale coherent structure enhances the small-scale thermalized fluctuations \cite{xu2017origins}.

One can see from Fig. \ref{fig:entropy} a steady growth of entropy with energy for both branches. However, the features of such growth differ. That is, the entropy increases along isogains for the $P_0^{+}$-branch but changes slowly for the $P_0^{-}$-branch. The change of entropy has a physical meaning, so the latter fact in connection with the DS energy quantization could be decisive for the single DS stability and, in general, its ability to self-emergence (``self-start''). 

\begin{figure}
    \centering
    \includegraphics[width=0.8\linewidth]{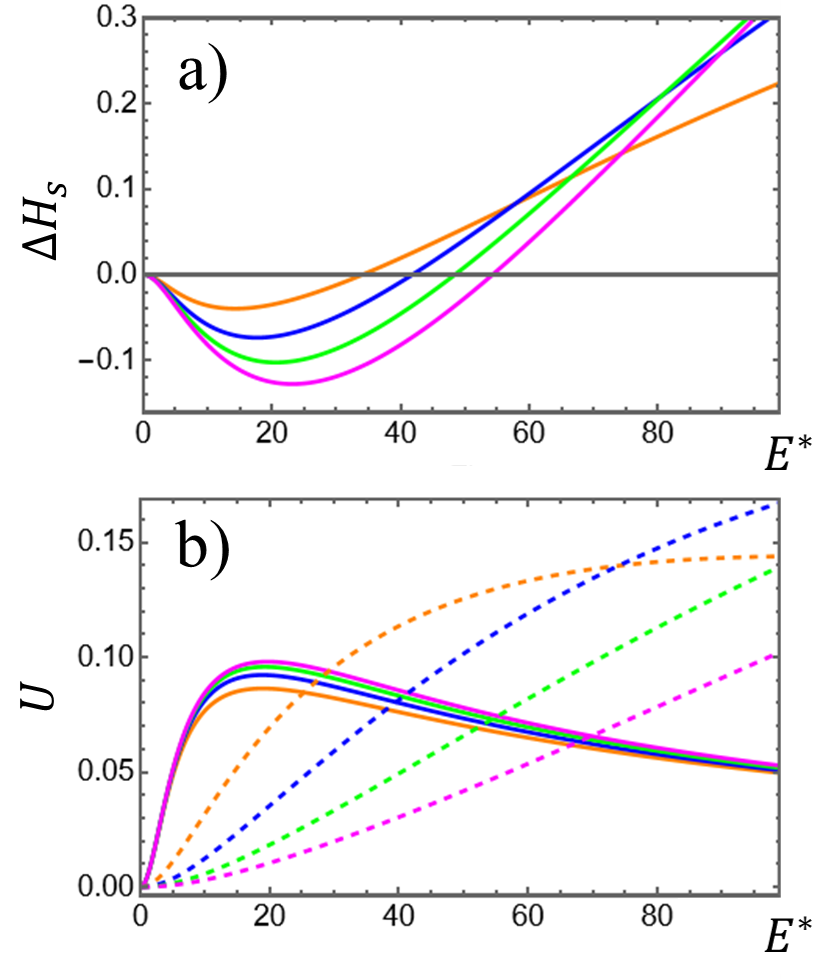}
    \caption{(a) The entropy differences $\Delta H_s$ between the multipulse $P_0^{-}$-complexes (orange -- two, blue -- three, green -- four, and magenta -- five pulses) and their single pulse $P_0^{+}$-counterparts with the same $\Sigma$ in dependence on $E^*$ (see Fig. \ref{fig:q}, b). (b) The internal energy $U$ for the multipulse complexes as in (a) (dashed curves) and their $P_0^{+}$-counterparts (solid curves).}
    \label{fig:ental}
\end{figure}

The differences $\Delta H_s$ between the additive entropy of $P_0^{-}$-complexes and their $P_0^{+}$-counterparts are shown in Fig. \ref{fig:ental}, a. The additivity means adding the individual entropy of each pulse under the assumption that they do not interact. One can see that $\Delta H_s$ becomes positive after $E^*\approx$40. The $\Delta H_s$ crossing zero (by analogy with the ``Maxwell point'') refers to conditions when single-pulse and multiple-DS states have equal ``thermodynamic-like potential'' (entropy or internal energy), indicating a balance point where one state can switch to another. That is, a multi-pulse state is thermodynamically preferable for larger energies. That correlates with the decrease of the order parameter in Fig. \ref{fig:fig6}, a, and is close to curves 1,2 in Fig. \ref{fig:fig1}, marking the transition to multiple pulses. These curves can be interpreted by analogy with the spinodal boundaries at which a single DS state can spontaneously become unstable, splitting into multiple pulses. 
%

\subsection{DS ``temperature''}\label{TETA}

\begin{figure}
    \centering
    \includegraphics[width=0.8\linewidth]{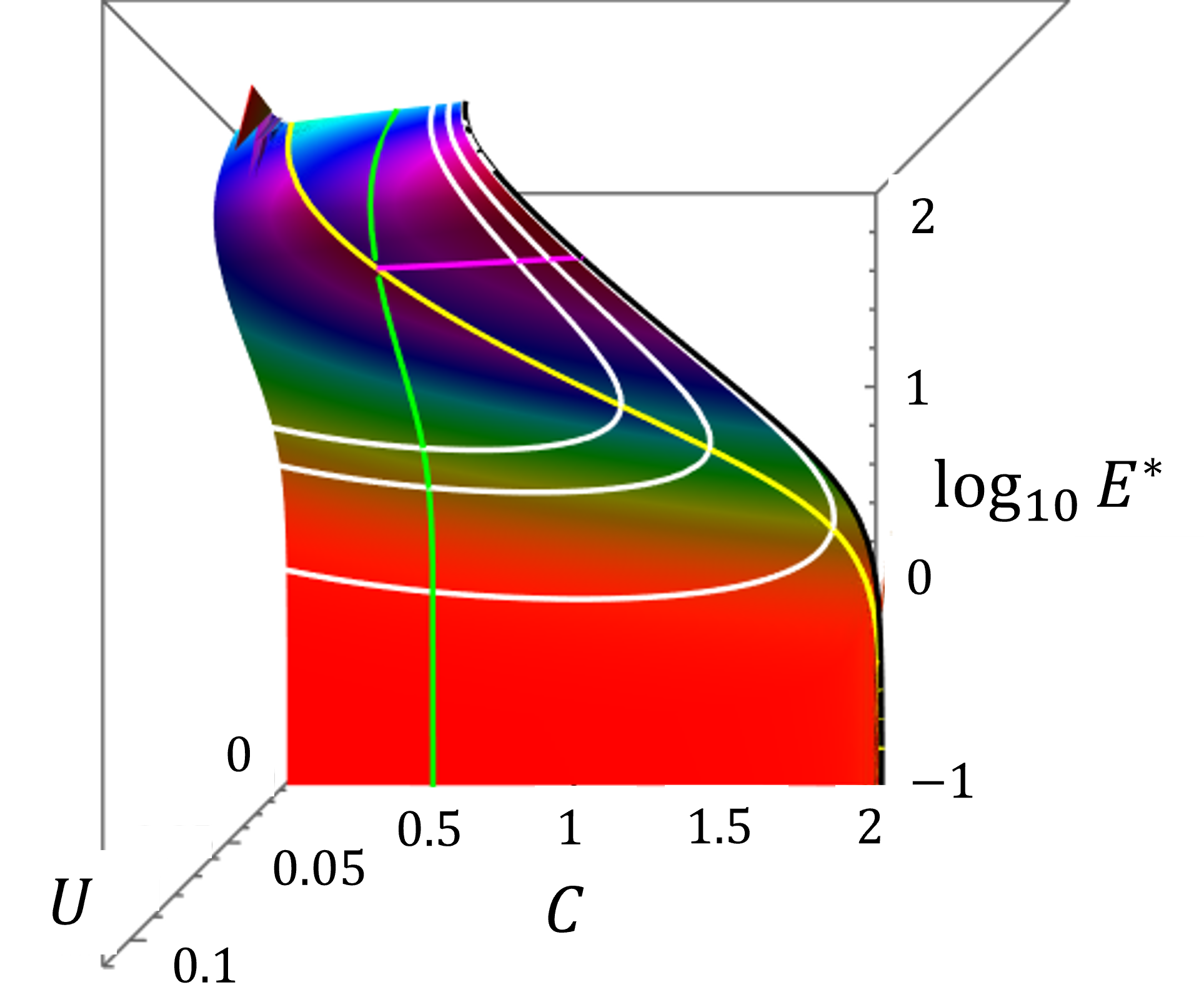}
    \caption{The DS internal energy $U$ versus $C$ and the logarithm of $E^*$ for the DS solutions $P_0^{+}$ (right side of the surface) and $P_0^{-}$ (left side of the surface). The curves correspond to those in Fig. \ref{fig:entropy}.}
    \label{fig:entalpy}
\end{figure}

Considering the internal energy $U$ (9) could provide deeper insight into the DS thermodynamic properties. Fig. \ref{fig:entalpy} shows the dependence of $U$ on $C$ and $E^*$ parameters. This Figure and Fig. \ref{fig:ental}, b (solid curves) demonstrate a change of the $\frac{\partial U}{\partial E^*}$-sign with the energy growth. That is, the internal energy begins to decrease with energy starting from some level of the latter. Since the entropy \textit{rises} with energy monotonically, a parallel \textit{decrease} of the internal energy means that the thermodynamic temperature $\Theta^{-1}=\frac{\partial H}{\partial U}$ becomes negative \cite{ramsey1956thermodynamics,marques2023observation}. This phenomenon can be interpreted as a consequence of the bounded DS spectrum within which there are intensive energy flows, which are induced by the strong chirp $\psi \gg 1$ \cite{ankiewicz2008dissipative,Sorokin17}, so it shares some conceptual similarities with turbulence (see Section \ref{th})\footnote{For DS considered in this work, there is a direct energy cascade from larger scales to smaller ones. The spectral dissipation and nonlinearity play a decisive role in forming such a cascade: energy injection is maximal at $\omega=0$ and dissipates with $|\omega|$-growth. Chirp, caused by nonlinearity, redistributes (scatters) energy from lower frequencies located around the DS center ($t=0$) to higher ones located close to the DS edges, where energy dissipates \cite{ankiewicz2008dissipative,Sorokin17,bale2008spectral}. Curiously, there is some analogy with 2D turbulence where the negative temperature corresponds to a discrete system of statistically independent vortices \cite{Kraichnan1980}. However, the direction of energy flux is reversed in our case.}.

\begin{figure}
    \centering
    \includegraphics[width=0.8\linewidth]{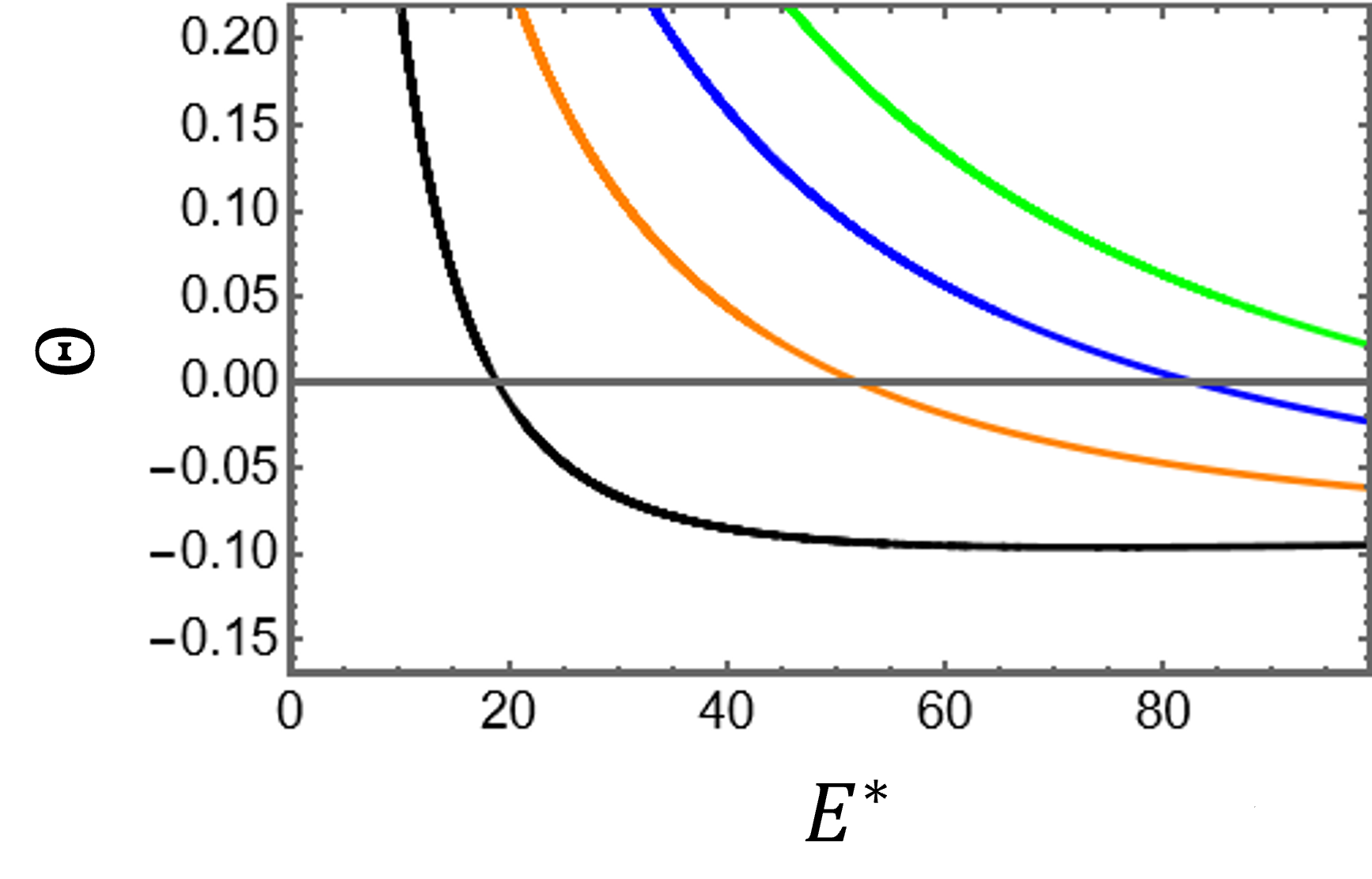}
    \caption{The temperature $\Theta$ evolution with $E^*$ for the ``quantized'' $P_0^{+}$-complexes in Fig. \ref{fig:q} having $P_0^{-}$-counterparts with two (black curve), three (orange), four (blue), and five (green) pulses.}
    \label{fig:temp}
\end{figure}

Fig. \ref{fig:temp} shows the temperature dependencies on $E^*$ for the different $P_0^{+}$-branches having two (black curve), three (orange), four (blue), and five (green) $P_0^{-}$-partners from Fig. \ref{fig:q}, b. The points of the $\Theta=0$--crossing with the transition to negative temperatures mean that the DS with larger energies becomes unstable, i.e., ``overheated,'' and such a non-equilibrium state tends to relax in a state with higher entropy and positive temperature, i.e., multiple $P_0^{-}$-pulses. The further energy growth causes a cascade of such transitions (Fig. \ref{fig:temp}). The real-time experimental observations of DS formation based on the dispersive Fourier transform technique \cite{goda2013dispersive} demonstrate the complex dynamics of this process, including a transition through an unstable ``overheated'' DS stage \cite{herink2016resolving,liu2018real,peng2018real,PhysRevResearch.2.013101}.\footnote{The high-entropy multi-DS state can form \textit{ab initio}, i.e., directly from the quantum noise after complex transient dynamics.}.

A transition to a negative temperature state indicates a reversal in the typical energetic ordering of states. Still, it does not necessarily signal a phase transition in the standard thermodynamical sense. For instance, free energy ($\mathcal{F}$) minimization cannot be considered a criterion of dynamic stability of DS due to the non-equilibrium character of a system (see Appendix \ref{secA3}).

\section{Discussion and Conclusion}\label{conc}

In this work, we considered a strongly chirped DS from the point of view of its energy scaling ability.  Such scalability can be termed a DSR, which, in principle, allows asymptotical energy growth due to the temporal DS broadening and spectral condensation. We describe this phenomenon based on the adiabatic approximation of the complex cubic-quintic nonlinear Ginzburg-Landau equation, which assumes a strong domination of the nondissipative factors, such as GDD and SPM, over the dissipative ones, such as spectral dissipation and SAM. 

The theory predicts three main manifestations of the transition to DSR regime: i) a transition from the DS squeezing to its asymptotical broadening, ii) a tendency of the DS spectral width to become constant, and iii) an appearance of a Lorentzian-like spike at the center of the DS spectrum. These phenomena are confirmed by our experiments with a Cr$^{2+}$:ZnS chirped-pulse oscillator.

The DS can be described entirely in the frameworks of a two-parametric ``master diagram'' composed from a manifold of ``isogains,'' i.e., curves of constant saturated net gain $\Sigma$, and divided into three domains: a region, where DS does not exist (region of vacuum instability $\Sigma<0$); a region of energy scalable DSs $P_0^{+}$, where a well-defined sub-region corresponds to DSR; and a region of energy unscalable DSs $P_0^{-}$.

In agreement with the experimental results, the theory predicts the Rayleigh-Jeans spectrum of DS, which has the frequency-bounded Lorentzian profile characterized by the width $\Xi$ and the cutoff frequency $\Delta$ (i.e., the experimental spectrum half-width). The Lorentzian spectral width $\Xi$, shortening with energy, and cutoff frequency $\Delta$, tending to a constant for DSR, define two correlation scales for DS, which can be interpreted as the scales defining the DS temporal width (``long scale'' $\Lambda=1/\Xi$) and the internal coherence scale (``short scale'' $l=\pi/\Delta$) appearing due to phase inhomogeneity. Their equality $\Xi=\Delta$ means the maximal DS ``fidelity'', i.e., the weakest frequency-dependence of the chirp, and, thereby, the maximal quality of the DS compression (approximately down to $\propto 1/\Delta$) by a compressor with the constant anomalous GDD. The ``fidelity'' curve on the master diagram defines the energies from which DSR starts.

The spectral cutoff inherent in the mechanism of DS formation is a source of the existence of two coherency scales. As a result, DS can be interpreted as an ensemble of coupled and localized ``quasi-particles'' (``microstates'') in a manner of the incoherent soliton description. The long scale $\Lambda$ defines the DS width (width of confining ``collective potential''), and the short scale $l$ defines a ``microstate.'' Hence, we conjecture that the DS spectrum could be interpreted as a probability distribution of ``microstates'' that makes possible the thermodynamic approach to the DS description.

We apply such an approach to study the limits of DS energy scalability. The numerical simulations, considering the quantum noise contribution in DS formation, demonstrate the tendency for multiple DS generation with energy increase. Following the experimental findings, this phenomenon limits the DS maximal energy scalability. We found that this process can be characterized by a probability distribution of multiple DSs, which broadens and shifts toward a larger number of pulses with energy growth. The model shows clear hysteresis and multi-stability: the DS solution can exist either as a single pulse or multiple pulses, depending on initial conditions or external perturbations (Fig. \ref{fig:fig5probability}). However, spontaneous transitions or decay are not seen when the regime is sustained. Such transitions appear during the initial stage of DS formation, which becomes metastable with the energy growth.

The transition to a metastable state is reflected by the increase in DS entropy. For DS, the entropy growth with energy is accompanied by a spectral condensation, i.e., a concentration of energy around the spectral center $\omega=0$ due to $\Xi \to 0$ with the parallel DS broadening. That differs DSR from BEC cardinally. Indeed, spectral condensation lowers the BEC entropy, but it increases the number of the DS ``microstates'' and entropy, resembling a transit to turbulence.

The generation of multiple DSs preventing unlimited energy scaling due to entropy growth is closely connected with the existence of two branches (``macrostates'') of DS, i.e., $P_0^{+}$ and $P_0^{-}$. The latter provides a possibility of the DS energy ``quantization'', i.e., the redistribution of the $P_0^{+}$-DS energy (scalable) between multiple $P_0^{-}$-DSs (energy unscalable). Since these branches have different thermodynamic properties, such redistribution becomes thermodynamically preferable after the dimensionless energy $E^*\approx$30, when the entropy difference between two $P_0^{-}$-pulses and a single $P_0^{+}$-pulse turns positive. 

However, the cascade of transitions to multiple pulsing could occur even earlier (at $E^*\approx$20) due to the convex behavior of the internal energy on $E^*$ (Fig. \ref{fig:entalpy}). This marks the transition into a region where the effective ``temperature'' $\Theta$ becomes \textit{negative}, indicating instability of the single DS solution. Our numerical calculations confirm these energy limits on the DS scalability.

We can conclude that the thermodynamic ideology can be applied to non-equilibrium systems such as DS, which provides deeper insight into the properties of semi-coherent dissipative structures ranging from nonlinear optics and weakly dissipative BEC to hydrodynamics and plasma physics. Even now, our results elucidate the road to energy scaling of DS in chirped-pulse oscillators as well as all-normal-dispersion fiber and waveguide lasers. This would be especially significant in developing high-power mid-infrared ultrashort pulse laser systems that allow progress in quantum sensors for metrology and environment monitoring, medicine and public security,  astrophysics, etc.     

\appendix
\section{Approximated DS solution of the complex cubic-quintic nonlinear Ginzburg-Landau equation}\label{secA1}
Here, we briefly describe the adiabatic approach to the solution of (\ref{eq:NGLE}) following Ref. \cite{podivilov2005heavily}. Let us start with the traditional soliton ansatz:
\begin{equation}\label{eq:a1}
a\! \left(z,t\right)=\sqrt{P\! \left(t\right)}\, {\mathrm e}^{\mathrm{i} \phi \left(t\right)-\mathrm{i} q z},
\end{equation}

\noindent which, after substitution to (\ref{eq:NGLE}) and separation into real and imaginary parts, gives a system
\begin{widetext}
\begin{gather}
    2 P\! \left(t\right) \left(\frac{d^{2}}{d t^{2}}P\! \left(t\right)\right) \beta -\left(\frac{d}{d t}P\! \left(t\right)\right)^{2} \beta +4 P\! \left(t\right) \left(\frac{d}{d t}P\! \left(t\right)\right) \Omega \! \left(t\right) \alpha +4 P\! \left(t\right)^{2} \left(-\beta  \Omega \! \left(t\right)^{2}+\alpha  \left(\frac{d}{d t}\Omega \! \left(t\right)\right)-\gamma  P\! \left(t\right)+q\right)=0, \label{eq:a2}\\
    2 P\! \left(t\right) \left(\frac{d^{2}}{d t^{2}}P\! \left(t\right)\right) \alpha -\left(\frac{d}{d t}P\! \left(t\right)\right)^{2} \alpha -4 P\! \left(t\right) \left(\frac{d}{d t}P\! \left(t\right)\right) \Omega \! \left(t\right) \beta -\nonumber\\-4 P\! \left(t\right)^{2} \left(\kappa  \zeta  P\! \left(t\right)^{2}+\Omega \! \left(t\right)^{2} \alpha +\left(\frac{d}{d t}\Omega \! \left(t\right)\right) \beta -\kappa  P\! \left(t\right)+\sigma \right)=0,\nonumber
    \end{gather}
\end{widetext}
\noindent where $\frac{d}{d t}\phi \! \left(t\right)=\Omega \! \left(t\right)$ is an instant frequency.

From the first of Eqs. (\ref{eq:a2}) with taking into account $\alpha \ll \beta$ and assuming the adiabatic change of the DS envelope $\frac{{d}^{2}}{d t^{2}}\left(\sqrt{P\! \left(t\right)}\right)\to 0$, one may obtain the expression for $P(t)$:
\begin{equation}\label{eq:a3}
    \gamma P\! \left(t\right)=q-\beta  \Omega \! \left(t\right)^{2}.
\end{equation}
\noindent The non-negativity of the power $P(t)$ results in the spectral cut-off condition from (\ref{eq:a3}):

\begin{equation}\label{eq:cutoff}
    \Omega(t)^2<\Delta^2=q/\beta.
\end{equation}

Taking into account the adiabatic condition and Eq. (\ref{eq:cutoff}), one may obtain from the second equation of (\ref{eq:a2}) the following equation:
\begin{widetext}
\begin{equation}\label{eq:a4}
    \beta\frac{d}{d t}\Omega \! \left(t\right)=\left[ \frac{2 \Omega \left(t\right)^{2}}{\Delta^{2}-\Omega \left(t\right)^{2}}-1\right]^{-1}\times\left[  \frac{\kappa  \beta  \left(\Omega \left(t\right)^{2}+\Delta^{2}\right) \left(\gamma +\zeta  \beta  \left(\Omega \left(t\right)^{2}-\Delta^{2}\right)\right)}{\gamma^{2}}-\sigma -\Omega \! \left(t\right)^{2} \alpha\right].
\end{equation}
\end{widetext}

Since the instant frequency should not be infinite, the first multiplier in (\ref{eq:a4}) should not equal zero. Excluding infinity can be provided by the regularization (factorization) procedure described in detail in \cite{kalashnikov2024dissipative}. As a result of regularization, we have two solutions for $\Delta^2$ or, under (\ref{eq:a3}), for the DS peak power $P_0^{\pm}=\beta \Delta^2/\gamma$: 
\begin{equation}\label{eq:power}
    \zeta P_{0}^{\pm}=\frac{3}{4 } \left[1-\frac{C}{2}\pm\sqrt{\left(1-\frac{C}{2}\right)^{2}-\Sigma}\right],
\end{equation}
\noindent where $C=\alpha \gamma/\beta \kappa$ and $\Sigma=4 \zeta  \sigma/\kappa$.

The regularized Eq. (\ref{eq:a4}) takes the form of
\begin{equation}\label{eq:a5}
    \frac{d}{d t}\Omega \! \left(t\right)=\frac{\beta  \kappa  \zeta}{3 \gamma^{2}}  \left(\Delta^{2}-\Omega \! \left(t\right)^{2}\right) \left(\Xi^{2}+\Omega \! \left(t\right)^{2}\right),
\end{equation}
\noindent where

\begin{equation}\label{eq:xi}
    \Xi^2=\frac{\gamma}{\zeta\beta}\left(C+1-\frac{5 \zeta P_{0} }{3}\right).
\end{equation}

Eq. (\ref{eq:a5}) can be solved implicitly:
\begin{gather}\label{eq:a6}
t=T\left[\frac{\Delta}{\Xi}\tan^{-1}\left(\frac{\Omega \left(t\right)}{\Xi}\right)+\mathrm{tan^{-1}}\! \left(\frac{\Omega \left(t\right)}{\Delta}\right)\right]^{2},\nonumber \\ 
T=\frac{3 \gamma^{2}}{\beta  \zeta  \kappa  \Delta  \left(\Delta^{2}+\Xi^{2}\right)},
\end{gather}
\noindent where $T$ plays a DS temporal width.

Eqs. (\ref{eq:a1},\ref{eq:a3}) lead to
\begin{equation}\label{a7}
    a\! \left(z,t\right)=\sqrt{\frac{-\beta  \Omega \! \left(t\right)^{2}+q}{\gamma}}\, {\mathrm e}^{i \left(\phi \left(t\right)-q z\right)}.
\end{equation}

Replacing $\Omega$ by $\omega$ and omitting the phase slip $q$ with the subsequent Fourier transformation give the DS spectral profile
\begin{equation}\label{eq:a8}
    e\! \left(\omega \right)=\int_{-\infty}^{\infty}\frac{{\mathrm e}^{-i \omega  t} \sqrt{\beta  \left(\Delta^{2}-\omega^{2}\right)}\, {\mathrm e}^{i \phi \left(t\right)}}{\sqrt{\gamma}}d t.
\end{equation}

Our assumption that DS is strongly chirped, i.e., its phase varies rapidly, allows applying the method of stationary phase \cite{bender2013advanced} to (\ref{eq:a8}). That results in the expression for a DS spectral amplitude:
\begin{gather}
    e\left(\omega \right) = \sqrt{\frac{6 \pi \gamma}{i \kappa \zeta}} \left(\Xi^{2}+\omega^{2}\right)^{-1/2}\times\nonumber\\
 \times \exp\left[ \frac{3 i}{2}\frac{\gamma^{2}}{\kappa \beta \zeta }\frac{\omega^2}{\left(\Xi^{2}+\omega^{2}\right) \left(\Delta^2-\omega^2\right)} \right] \mathcal{H}(\Delta^2-\omega^2), \label{eq:sa}
\end{gather}
\noindent a spectral power $p(\omega)=|e(\omega^2)|$, and an energy $E=\frac{1}{2\pi} \int_{-\infty }^{\infty }p(\omega)d\omega$.

\section{Autocorrelation function for a strongly chirped DS of the cubic nonlinear Ginzburg-Landau equation}\label{secA2}

Let us consider a version of (\ref{eq:NGLE}) with $\zeta=0$ (Haus's master equation \cite{902165}) describing a laser, which is passively mode-locked by a fast saturable absorber (such as, e.g., Kerr-lens mode-locking) without taking into account SAM saturation. When $\psi \gg 1$, one may apply the method of stationary phase for the calculation of spectral power of the exact DS solution $a(t)=\sqrt{P_0}\sech(t/T)^{1+i\psi}$ that gives for a spectral power:
\begin{equation}\label{eq:a9}
    p(\omega)=\frac{2\pi P_0}{\Delta^2}(1+\omega^2 T/\Delta)^{-1}\mathcal{H}(\Delta^2-\omega^2),
\end{equation}
\noindent where a spectral width (cut-off frequency) is defined by $T\Delta =\psi$. Then, the first-order autocorrelation function behaves as
\begin{equation}\label{eq:ac2}
R(\tau) = \frac{P_0 \Delta}{2 \sqrt{\theta}} \int_{-\infty}^{\infty} e^{- \frac{|t|}{\sqrt{\theta}}} \, \sinc\left( \Delta (\tau - t) \right) \, dt
\end{equation}
\noindent with two characteristic time scales $l=\pi/ \Delta$ and $\Lambda=\sqrt{\theta}=\sqrt{T/\Delta}$.
According to Ref. \cite{kalashnikov2009unified}, $\Delta^2 = 3\sigma C/\alpha (2-C)$, $E=6\beta \Delta/(1+C)\kappa$, $P_0=\beta \Delta^2/\gamma$, and $T=3\gamma/\kappa(1+C)\Delta$. The DSR condition is $C \to 2$, and the DS energy scaling is accompanied by the DS squeezing, spectral broadening, and peak power growth that differs from the DSR scenario considered above due to the absence of the SAM saturation. 

\section{DS spectral perturbations}\label{secAperturb}

Let us consider the perturbations of DS (\ref{eq:a9}) caused by a quintic nonlinearity in (\ref{eq:NGLE}). The adiabatic theory allows for developing a perturbation theory in the spectral domain \cite{akhmediev1997nonlinear,kalashnikov2011dissipative}. 

The equation for the Fourier image $f(\omega)$ of the small perturbation of $p(\omega)$ can be written as 

\begin{widetext}
\begin{equation}
   S(\omega) = \left( q-k(\omega) \right)f(\omega)+\frac{1}{\pi}\int_{-\infty }^{\infty }U(\omega-\omega^{\prime})f(\omega^{\prime})d\omega^{\prime} +\frac{1}{2\pi}\int_{-\infty }^{\infty }V(\omega-\omega^{\prime})f^*(\omega^{\prime})d\omega^{\prime},\label{eq:per1}
\end{equation}
\end{widetext}

\noindent where $k(\omega)=\beta \omega^2 + i\left( \sigma+\alpha \omega^2 \right)$ is the wavenumber and kernels $U$ and $V$ are the Fourier images of $|a(t)|^2$ and $a(t)^2$, respectively. The source term $S(\omega)$ is a Fourier image of $-i\kappa \zeta |a|^2 a$.

\begin{figure}[h]
    \centering
    \includegraphics[width=0.75\linewidth]{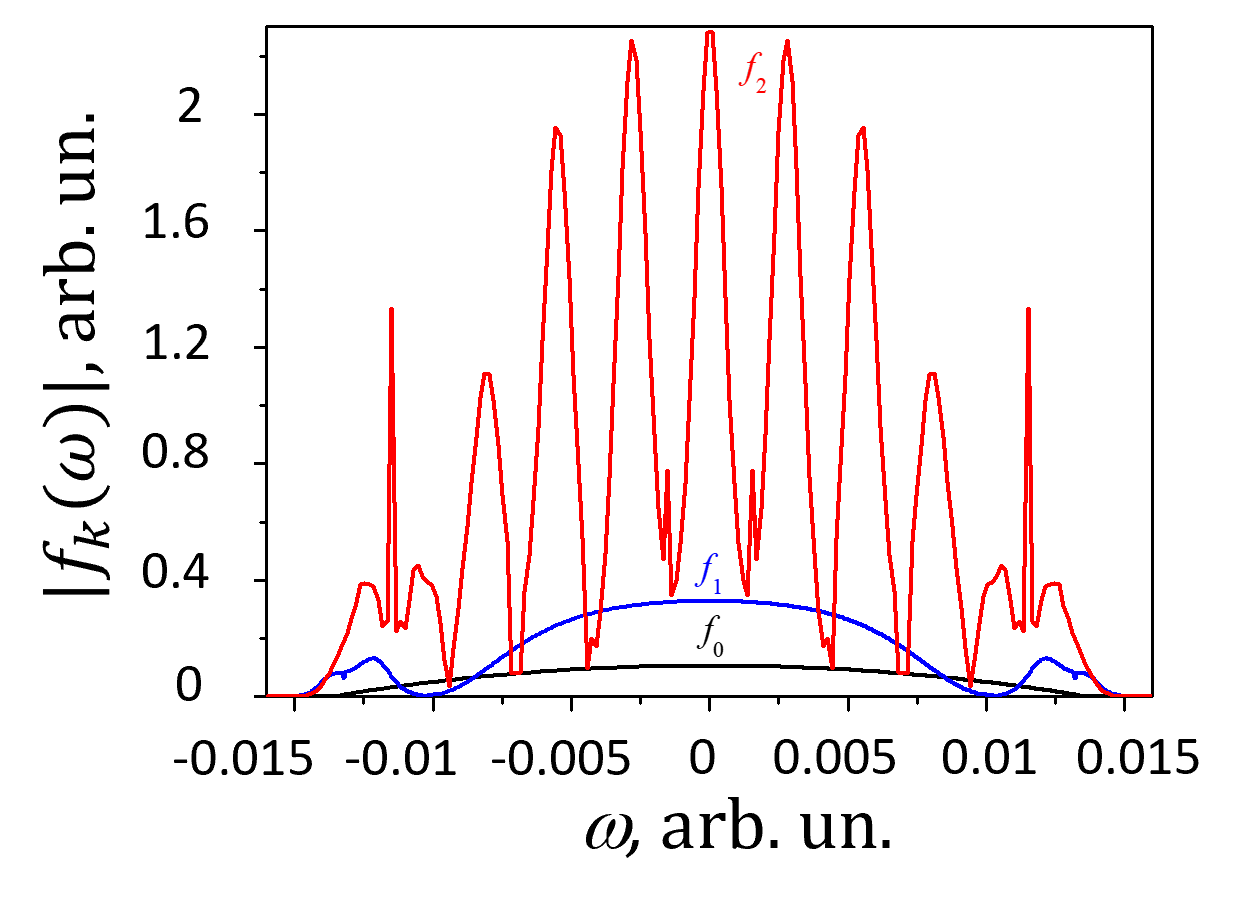}
    \caption{The absolute values of the successive terms $f_0$ (black), $f_1$ (blue), and $f_2$ (red) in the Neumann series.}
    \label{fig:perturbed}
\end{figure}

The assumption of a phase matching between the soliton and its perturbation allows $U=V$ and Eq. (\ref{eq:per1}) can be solved in the form of Neumann's series for $f_n$:

\begin{widetext}
\begin{equation}
    f_n(\omega)=\frac{S(\omega)}{q-k(\omega)}-\frac{3}{\pi \left( q-k(\omega) \right)}\int_{-\infty }^{\infty }U(\omega-\omega^{\prime})f_{n-1}(\omega^{\prime})d\omega^{\prime},\label{eq:per2}
\end{equation}
\end{widetext}
\noindent where $f_0=S(\omega)/[q-k(\omega)]$.
The first three $f_n(\omega)$ profiles are shown in Fig. \ref{fig:perturbed}, which demonstrate the appearance of a quasi-periodical internal structure.

\section{DS thermodynamics}\label{secA3}

Let us interpret (\ref{eq:spectrum}) as an unnormalized Rayleigh-Jeans probability distribution where the micro-states are the soliton quasiparticles in the sense of Refs. \cite{akhmediev2000multi,picozzi2007towards,picozzi2009thermalization,picozzi2014optical}. The distribution (\ref{eq:spectrum}) must be normalized so that $\int_{-\Delta}^{\Delta}p(\omega)d\omega=1$, leading to a normalized distribution (we will omit the star-superscripts for the dimensionless values):
\begin{gather}\label{eq:dist}
   \tilde p(\omega)= p(\omega)/Z = \frac{\Xi}{2\left( \Xi^2+\omega^2 \right) \tan^{-1}(\frac{\Delta}{\Xi})},\\ \nonumber Z=\int_{-\infty }^{\infty }p(\omega)d\omega=12\pi \Xi^{-1}\tan^{-1}(\frac{\Delta}{\Xi}).
\end{gather}

\noindent Then the entropy can be expressed in the form of

\begin{widetext}
\begin{equation}\label{eq:h}
    H=-\int_{-\Delta}^{\Delta} \tilde p(\omega)\ln(\tilde p(\omega)) d\omega= \ln\left[ 2 \Xi \tan^{-1}\left( \dfrac{\Delta}{\Xi} \right) \right] + \dfrac{1}{\tan^{-1}\left( \dfrac{\Delta}{\Xi} \right)} \int_{0}^{\Delta/\Xi} \dfrac{\ln(1 + x^2)}{1 + x^2} \, dx.
\end{equation}
\end{widetext}

\noindent Now, we need to re-scale the expression for entropy to make it positive and exclude an infinity $\lim_{\Xi \to 0}H=-\infty$. The shifted expression for entropy reads as:
\begin{equation}\label{eq:hc}
    H_s=\dfrac{1}{\tan^{-1}\left( \dfrac{\Delta}{\Xi} \right)} \int_{0}^{\Delta/\Xi} \dfrac{\ln(1 + x^2)}{1 + x^2} \, dx.  
\end{equation}


\noindent Using the definition of the internal energy $U=\left( 1/2 \right)\int_{-\Delta}^{\Delta}\tilde p(\omega)\omega^2d\omega$ \cite{wu2019thermodynamic} leads to
\begin{equation} \label{eq:a10}
   U=\frac{\Xi}{2}\left[ \frac{\Delta}{\tan^{-1}\left( \frac{\Delta}{\Xi} \right)} -\Xi\right].
\end{equation}

Then, the temperature $\Theta=\left( \frac{\partial H_s}{\partial U} \right)^{-1}$ can be defined from Eqs. (\ref{eq:hc},\ref{eq:a10}) and calculated numerically. The commented Mathematica code is available for downloading \cite{code}.  

\begin{figure}[h]
    \centering
    \includegraphics[width=0.75\linewidth]{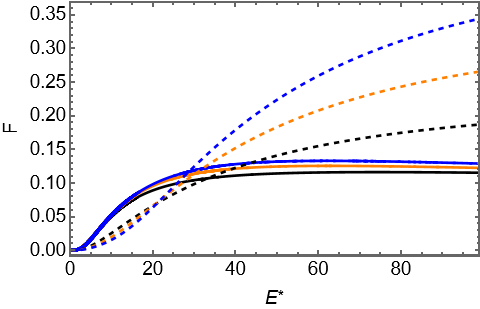}
    \caption{Free energy $\mathcal{F}$ for the multipulse complexes (two-pulse - black dashed curve, three - orange, and four - blue) and their $P_0^{+}$-counterparts (corresponding solid curves).}
    \label{fig:freeenergy}
\end{figure}

The standard relation for the free energy:

\begin{equation} \label{eq:a11}
   \mathcal{F}=U-\Theta H_s
\end{equation}

\noindent allows its calculation from the previously obtained thermodynamic quantities (Fig. \ref{fig:freeenergy}) \cite{code}. However, as a result of the non-equilibrium character of a system, the stable state does not correspond to the criterion of free energy minimization. Nevertheless, the equality of free energies could be considered by analogy with the Maxwell points, manifesting a transition between single- and multiple-DS states (Fig. \ref{fig:freeenergy}). Also, the tendency to such transition is demonstrated by the concave behavior of the $\mathcal{F}$ dependence on $E^*$ for a single DS (solid curves).


\begin{acknowledgements}
  This work was supported by Norges Forskningsr{\aa}d (\#303347 (UNLOCK), \#326241 (Lammo-3D), \#326503 (MIR)), and ATLA Lasers AS. We acknowledge using the IDUN NTNU scientific cluster \cite{sjalander}.
\end{acknowledgements}

\bibliography{bibliography}
\end{document}